\documentclass[aps,prd,preprint,groupedaddress,nofootinbib]{revtex4}
\usepackage{graphicx}
\usepackage{amsmath,amssymb}
\usepackage{bm}
\usepackage[caption=false]{subfig}
\usepackage{color}
\usepackage{floatrow}
\usepackage{setspace}
\usepackage[x11names]{xcolor}
\usepackage[colorlinks,linkcolor=purple,citecolor=teal]{hyperref}

\def\Lag{\mathcal{L}}
\def\p{\partial}

\def\=:{=\hspace{-.7em}\raisebox{1.1ex}{.}\hspace{.1em}\raisebox{-0.2ex}{.}}
\newcommand{\tr}{{\rm tr}\,}
\newcommand{\beq}{\begin{eqnarray}}
\newcommand{\eeq}{\end{eqnarray}}
\newcommand{\non}{\nonumber\\}

\newcommand{\bphi}{\boldsymbol{\phi}}
\newcommand{\U}{{\rm U}}
\newcommand{\SU}{{\rm SU}}
\newcommand{\Og}{{\rm O}}

\renewcommand{\i}{\mathrm{i}}
\renewcommand{\d}{\mathrm{d}}

\newtheorem{definition}{Definition}
\newtheorem{conjecture}{Conjecture}
\begin{document}

\title{Linked vortices as baryons in the miscible BEC-Skyrme model}

\author{Sven Bjarke Gudnason}
\email{gudnason(at)henu.edu.cn}
\affiliation{Institute of Contemporary Mathematics, School of
  Mathematics and Statistics, Henan University, Kaifeng, Henan 475004,
  P.~R.~China
}
\author{Muneto Nitta}
\email{nitta(at)phys-h.keio.ac.jp}
\affiliation{Department of Physics, and Research and Education Center for Natural 
Sciences, Keio University, Hiyoshi 4-1-1, Yokohama, Kanagawa 223-8521, Japan
}
\date{\today}
\begin{abstract}
We introduce a variation of the Bose-Einstein condensate(BEC)-Skyrme
model, with an altered potential for miscible BECs that gives rise to
two physical vortex strings. 
In the ground state of each topological sector, the vortices are
linked exactly $B$ times, due to a recently formulated theorem, with
$B$ being the baryon number of the solution.
The model also possesses metastable states, where the vortices are
degenerate and do not lend the interpretation of the baryon number
as the linking number of the vortices.

\end{abstract}
%\pacs{}
%\keywords{}
\maketitle

\tableofcontents

\section{Introduction}

More than a century ago, Lord Kelvin imagined that atoms were made of
knotted vortices \cite{Thomson:1869}, but this idea has not been
successful so far.
However, recently we have made a connection between not knots, but
links of vortices and Skyrmions in the Skyrme model 
\cite{Gudnason:2020luj}.
Skyrmions are solitons of the texture type in 3-dimensional space and
possess a topological degree $B$ as they are mapped to a 3-sphere,
being the target space or isospin space
\cite{Skyrme:1961vq,Skyrme:1962vh}.
In the large-$N$ limit of QCD, the baryon is identified with the
Skyrmion \cite{Witten:1983tw,Witten:1983tx}, thus providing a
solitonic approach to nuclear physics, see
e.g.~Refs.~\cite{Battye:2006na,Battye:2009ad,Baldino:2017mqq}. 
The Skyrmion may also be realized in two-component Bose-Einstein
condensates (BECs)
\cite{Ruostekoski:2001fc,Battye:2001ec,Khawaja:2001,Khawaja:2001zz,Savage:2003hh,Ruostekoski:2004pj,Wuster:2005,Herbut:2006sw,Tokuno:2009,Kawakami:2012zw,Nitta:2012hy},
see Ref.~\cite{Kasamatsu:2005} for a review.
In Refs.~\cite{Gudnason:2014gla,Gudnason:2014hsa,Gudnason:2014jga} we
introduced a potential inspired by two-component BECs, $V\sim
M^2|\phi_1|^2|\phi_2|^2$,  which deforms the Skyrmions into a twisted
vortex ring (or vorton) by explicitly breaking the SU(2) isospin
symmetry normally possessed by Skyrmion solutions.
Due to the nonlinear sigma model constraint,
$|\phi_1|^2+|\phi_2|^2=1$, the center of the vortex ring in one
component confines the other component; it is thus a global analog of
Witten's superconducting cosmic string \cite{Witten:1984eb}.
Similar vortex rings can also be obtained in a different asymmetric
potential, $V\sim M^2|\phi_1|^2$ \cite{Gudnason:2016yix}.
In addition to a vortex ring, the BEC-inspired potential also allows
for a domain wall
\cite{Gudnason:2014gla,Gudnason:2014hsa,Gudnason:2014jga} into which
the vortex ring can be absorbed, 
creating a vortex handle -- or rather a link of a handle and a dual
handle from both sides of the domain wall \cite{Gudnason:2018oyx}.
We would like to stress that the Skyrmions we study here are closer to
those describing nuclei within the Skyrme model, than to the Skyrmions
in BEC, because of the derivative part of the Lagrangian.
The potential and the terminology is borrowed from BECs and hence
similar behavior is expected.

The connection between 
links of vortices and Skyrmions \cite{Gudnason:2020luj}
is made by means of a theorem stating that there exists
a projection of the Skyrmion field onto a 2-sphere (as opposed to the
3-sphere that is the target space), which has the properties that two
distinct regular points are linked $B$ times in real physical
3-space, with $B$ being the baryon number or topological degree of the
Skyrmion.
Taking a natural Ansatz for the Skyrmion, like the one giving rise to
a $B$-twisted vortex ring
\cite{Gudnason:2014gla,Gudnason:2014hsa,Gudnason:2014jga}, produces
a ``physical'' vortex with winding number one and a ``vacuum''
vortex with winding number $B$.
The ``vacuum'' vortex is not physical in the sense that it strays off
to infinity in such a diluted form that the total energy is finite. 
As mentioned, the catch is that the points under the projection have
to be \emph{regular} points, which is not always the case.
In Ref.~\cite{Gudnason:2020luj} we have circumvented the issue by
introducing a rotation of the 2-sphere, hence making it possible to
find points which are regular and hence giving rise to nondegenerate
vortices that thus provide the linking number $Q=B$.
One may consider such a rotation a bit arbitrary and thus wonder about
the physical implication thereof.

In this paper, we modify the BEC-inspired potential studied in
Refs.~\cite{Gudnason:2014gla,Gudnason:2014hsa,Gudnason:2014jga,Gudnason:2018oyx}
by flipping the sign of the potential (which is allowed in a nonlinear
sigma model) and this is suitable for the class of BECs called
miscible two-component BECs, 
for which both the components $\phi_1$ and $\phi_2$ develop 
vacuum expectation values (VEVs) in the vacua.  
The implications thereof is a rather crucial change in the vacuum
structure and indeed the model only possesses a unique vacuum state
after the altercation, which we dub the miscible BEC-Skyrme model.
This case admits two kinds of vortices having windings in the
$\phi_1$ and $\phi_2$ components as in the case of the miscible BEC  
\cite{Kasamatsu:2005,Eto:2011wp,Kasamatsu:2015cia}.
As a consequence, the above-mentioned ``vacuum'' vortex is transformed
into a second ``physical'' vortex, thus realizing, in a physical way,
the two linked vortices proposed in Ref.~\cite{Gudnason:2020luj}.

Moreover there is the issue of regular points leading to nondegenerate
links of vortices versus singular points leading to degenerate
vortices in the framework of Ref.~\cite{Gudnason:2020luj}.
This notion takes a very physical form in the presence of the miscible
BEC-inspired potential, as the degenerate vortex links generally give
rise to higher-energy (metastable) states and the nondegenerate vortex
links yield stable (ground) states. 

This paper is organized as follows.
In Sec.~\ref{sec:model} we introduce the miscible BEC-Skyrme model and
discuss the vacuum structure of the model.
In Sec.~\ref{sec:lnkvtx} we define the numerical observables, explain
the numerical method, define the initial conditions used throughout
the paper and discuss the numerical results. 
Finally, we conclude the paper with a discussion in
Sec.~\ref{sec:discussion}.

\section{The miscible BEC-Skyrme model}\label{sec:model}

We consider the generalized Skyrme model which contains the kinetic
term, the Skyrme term \cite{Skyrme:1961vq,Skyrme:1962vh} and the
BPS-Skyrme (Bogomol'nyi-Prasad-Sommerfield-Skyrme) term
\cite{Jackson:1985yz,Adam:2010fg,Adam:2010ds}
  \footnote{The BPS-Skyrme
  term is named after the BPS-Skyrme model of
  Refs.~\cite{Adam:2010fg,Adam:2010ds}, because it possesses a
  saturable energy bound (BPS bound); this model consists only of the
  BPS-Skyrme term and a potential.  }
as well as a potential 
\begin{align}
  \Lag &= \Lag_2 + c_4\Lag_4 + c_6\Lag_6 - V,\label{eq:Lag}\\
  \Lag_2 &= -\frac12\p_\mu\bphi^\dag\p^\mu\bphi,\\
  \Lag_4 &= \frac18(\p_\mu\bphi^\dag\p_\nu\bphi)(\p^{[\mu}\bphi^\dag\p^{\nu]}\bphi)
    + \frac18(\p_\mu\bphi^\dag\sigma^2\p_\nu\bphi)(\p^{[\mu}\bphi^\dag\sigma^2\p^{\nu]}\bphi) \non
    &= -\frac14(\p_\mu\bphi^\dag\p^\mu\bphi)^2
    +\frac{1}{16}(\p_\mu\bphi^\dag\p_\nu\bphi + \p_\nu\bphi^\dag\p_\mu\bphi)^2,\label{eq:L4}\\
  \Lag_6 &= \frac14(\epsilon^{\mu\nu\rho\sigma}\bphi^\dag\p_\nu\bphi\p_\rho\bphi^\dag\p_\sigma\bphi)^2,\label{eq:L6}
\end{align}
where $\sigma^a$ are the Pauli matrices,
$\bphi\equiv(\phi_1(x),\phi_2(x))^{\rm T}$ is a complex two-vector 
of scalar fields which obey the nonlinear sigma model constraint
$\bphi^\dag\bphi=|\phi_1|^2+|\phi_2|^2$, the spacetime indices
$\mu,\nu,\rho,\sigma$ run over 0 through 3 and the flat Minkowski
metric is taken to be of the mostly positive signature.
The relation between the complex 2-vector of scalar fields, $\bphi$,
and the chiral Lagrangian field usually used in Skyrme-type models is
given by 
\beq
U =
\begin{pmatrix}
  \bphi & -\i\sigma^2\bar{\bphi}
\end{pmatrix}
=
\begin{pmatrix}
  \phi_1 & -\bar{\phi}_2\\
  \phi_2 & \bar{\phi}_1
\end{pmatrix},
\eeq
which translates the constraint $\bphi^\dag\bphi=\det U=1$ into the
usual one in terms of $U$.

The Skyrme term \eqref{eq:L4} consists of a particular combination of
two operators in the chiral Lagrangian, where the coefficients are
tuned so that the fourth-order time derivative of $U$ cancels out, see
for example Ref.~\cite{Gudnason:2017opo} for details.
The BPS-Skyrme term \eqref{eq:L6}, likewise is a particular
combination of three operators in the chiral Lagrangian at order
$p^6$ (six derivatives), where again the coefficients are tuned so
that there is only a second-order time derivative of $U$, but no
fourth order or sixth order ones (see Ref.~\cite{Gudnason:2017opo}).

With the potential turned off, the target space manifold is given by
$\Og(4)/\Og(3)\simeq\SU(2)\simeq S^3$ which is a unit 3-sphere. This
is due to the model without a potential possessing $\Og(4)$ symmetry,
which however would spontaneously break to $\Og(3)$ by the requirement
of finite energy.
The maps with finite energy necessarily have vanishing derivatives at
spatial infinity, which effectively point compactifies $\mathbb{R}^3$
to $\mathbb{R}^3\cup\{\infty\}\simeq S^3$.
Hence, topological solitons are supported with topological degree
$B\in\pi_3(S^3)=\mathbb{Z}$, where $B$ is given by
\beq
B = \frac{1}{4\pi^2}\int \d^3x\;\mathcal{B}, \qquad
\mathcal{B} \equiv \epsilon^{ijk}
\bphi^\dag\p_i\bphi\p_j\bphi^\dag\p_k\bphi,
\eeq
where $i,j,k=1,2,3$ are spatial indices.

In this paper we will augment the model with the miscible BEC-inspired
potential 
\beq
V = \frac18M^2\left[(\bphi^\dag\sigma^3\bphi)^2 - 1\right]
  = -\frac12M^2|\phi_1|^2|\phi_2|^2,
\label{eq:V}
\eeq
which is the same as the BEC-inspired potential used in
Refs.~\cite{Gudnason:2014gla,Gudnason:2014hsa,Gudnason:2014jga}, however
with $M^2\to -M^2$.
The previous and current ones are called immiscible and miscible,
respectively, in two-component BECs.
This change of the sign of the potential has crucial influence on the
vacuum structure since the vacuum now breaks the continuous symmetry
of the model completely; therefore both complex scalars gain VEVs and
describe a miscible BEC phase.

Turning on the potential \eqref{eq:V} explicitly breaks the symmetry
from $\Og(4)$ to
\beq
G = \U(1)\times\Og(2)
\simeq \U(1)_0\times[\U(1)_3\rtimes(\mathbb{Z}_2)_{1,2}],
\eeq
where the group actions defined by the above symmetries are given by
\begin{align}
  \U(1)_0\ \ &: \qquad
  \bphi\to e^{\i\omega}\bphi,\\
  \U(1)_3\ \ &: \qquad
  \bphi\to e^{\i\eta\sigma^3}\bphi,\\
  (\mathbb{Z}_2)_{1,2}\ \ &: \qquad
  \bphi\to e^{\i\frac{\pi}{2} (a \sigma^{1}+b \sigma^{2})}\bphi,  
\end{align}
with $a^2 + b^2 =1$.
The $(\mathbb{Z}_2)_{1,2}$ exchanges $\phi_1$ and $\phi_2$ as 
\begin{align}
 (\phi_1 , \phi_2)^{\rm T} 
 \to (e^{\i \gamma} \phi_2,  e^{-\i \gamma}\phi_1)^{\rm T},
  \label{eq:Z2}
\end{align}
with $e^{\i \gamma} = b+\i a$.
$\U(1)_3$ is acting on $(\mathbb{Z}_2)_{1,2}$ transforming $\gamma$
such that they define a semidirect product which we have denoted by
$\rtimes$. 

In stark contrast to the immiscible BEC-inspired potential, the miscible
version accommodates only a connected vacuum state
\beq
\bphi_{\rm vac} = \frac{1}{\sqrt{2}}(e^{\i\alpha},e^{\i\beta})^{\rm T},
\label{eq:vacuum}
\eeq
spontaneously breaking the symmetry $G$ into 
a ${\mathbb Z}_2$ subgroup (\ref{eq:Z2}) with 
\beq
 \gamma = \alpha - \beta ,
\eeq
which we denote by $({\mathbb Z}_2)_{\alpha-\beta}$.
The target space is thus given by
\beq
\mathcal{M} = G/H = [\U(1)_0\times \Og(2)]/
({\mathbb Z}_2)_{\alpha-\beta} = U(1)_1 \times U(1)_2,
\eeq
which does not allow for domain walls because of 
\beq
\pi_0(\mathcal{M}) = \mathbf{1},
\eeq
but accommodates two types of vortices supported by 
\beq
\pi_1(\mathcal{M}) = (\mathbb{Z})_1\times(\mathbb{Z})_2
\eeq
as the case of miscible two-component BECs 
\cite{Kasamatsu:2005,Eto:2011wp,Kasamatsu:2015cia}.

The absence of the domain wall 
is in contrast to the case of immiscible BEC-Skyrme model
with the spontaneously breaking of $(\mathbb{Z}_2)_{1,2}$  
admitting a domain wall. 
The immiscible BEC-Skyrme model was shown to contain vortex rings in
Ref.~\cite{Gudnason:2018oyx} and when pushed toward a domain wall --
present for the immiscible BEC-inspired potential -- the vortex ring
turned into a handle (in say $\phi_1$) on the domain wall, albeit
creating a second dual handle (in $\phi_2$).
The two seemingly different vortices were always present at the same
time and hence topologically the immiscible BEC-inspired potential of
Ref.~\cite{Gudnason:2018oyx} gives rise to a symmetry breaking
supporting only one topological number for the vortices.
In principle, in this model we can have two independent numbers of
vortices. 

The parameter space of the model with $c_4$, $c_6$, and $M$ is rather
large and for this reason we will choose only two model points
\begin{align}
&2+4\ {\rm model}\;:\qquad
c_4=1,\quad c_6=0,\\
&2+6\ {\rm model}\;:\qquad
c_4=0,\quad c_6=1,
\end{align}
which disentangles the effects of the Skyrme term ($c_4\neq 0$) and
the BPS-Skyrme term ($c_6\neq 0$).

\section{Skyrmions as linked vortices}\label{sec:lnkvtx}

\subsection{Local observables}

In order to observe the linked vortices in Skyrmions we need some
observables to identify the vortices.
We will consider the baryon charge density $\mathcal{B}$, the static
energy density $\mathcal{E}=-\Lag$, the potential energy (density) $V$
as well as a vorticity density.
In order to define the latter, we will construct a term inspired by
BPS vortices in the Abelian-Higgs model at critical coupling.
Starting with the BPS equation
\begin{align}
F_{12} &= e^2(|\phi|^2 - v^2), \\
D_{\bar{z}}\phi &= 0, \label{eq:selfdual_vtx}
\end{align}
and aiming at constructing an expression for the topological vortex
charge
\beq
Q = \frac{1}{2\pi}\int \d^2x\; F_{12},
\eeq
we can solve for the gauge field from Eq.~\eqref{eq:selfdual_vtx}
obtaining
\beq
A_{\bar{z}} = -\i\p_{\bar{z}}\log \phi,
\eeq
from which we can readily construct $F_{12}$:
\beq
F_{12} = -2\i F_{z\bar{z}}
= -2\p_z\p_{\bar{z}}\log|\phi|^2
= -\frac12(\p_1^2+\p_2^2)\log|\phi|^2,
\eeq
which is an expression for the topological vortex charge density for a
vortex pointing in the $x^3$ direction.
A simple extension of the charge density to cover a vortex pointing in
any direction in $\mathbb{R}^3$ can thus be made as
\beq
\mathcal{Q}_{1,2} = \frac12\sum_i\epsilon_{ijk}F_{jk}
= -(\p_1^2+\p_2^2+\p_3^2)\log|\phi_{1,2}|^2,
\label{eq:Q12}
\eeq
where we have added a label for the two complex fields in our model. 

For definiteness, we will define the total energy by the integral
expression
\beq
E_B = \int\d^3x\; \mathcal{E} = -\int\d^3{x}\;\Lag,
\eeq
where the subscript $B$ labels the topological sector at hand. 

\subsection{Numerical method}

In order to explore the model, we need to find numerical solutions to
the equations of motion, which we will solve by using the arrested
Newton flow method \cite{Gudnason:2020arj} for a cubic $100^3$
lattice, where the spatial derivatives are discretized using a
standard finite difference scheme with a fourth-order 5-point
stencil. 
The arrested Newton flow is updated in the ``time direction'' using a
fourth-order Runge-Kutta method.

\subsection{Initial conditions}

In this paper, the linked vortices will naturally be physical due to
the potential \eqref{eq:V}.
Nevertheless, it is not \emph{a priori} guaranteed that the linked
vortices are not degenerate, obscuring the interpretation that the
linking number is equal to the topological degree of the Skyrmions, as
proved by the theorem given in Ref.~\cite{Gudnason:2020luj}.
One could expect that increasing the coupling of the potential $M^2$,
would eventually force the linked vortices to be nondegenerate.
This will be investigated numerically in the next subsection. 

As initial conditions we use both the rational map approximations of
Ref.~\cite{Houghton:1997kg}, whose linked vortices were interpreted
and studied in Ref.~\cite{Gudnason:2020luj}, and composite Skyrmions
constructed from lower-charge solutions using the symmetric product
Ansatz
\beq
U^{\rm prod} = \frac{U_1 U_2 + U_2 U_1}{\det(U_1 U_2 + U_2 U_1)}.
\eeq
As already mentioned in the introduction, the linked vortices in
Ref.~\cite{Gudnason:2020luj} naturally contain a vacuum vortex, which
is not particularly physical, because it extends to infinity in
$\mathbb{R}^3$ although its flux is diluted such that the energy is
finite.
In Ref.~\cite{Gudnason:2020luj} we introduced a rotation of the
2-sphere after applying the Hopf map to a Skyrmion map, which
effectively eliminated the degeneracy of the linked vortices, allowing
for the interpretation of the linking number as the topological charge
or baryon number of the Skyrmions.

In this paper, we do not need to perform such a rotation of the
2-sphere, but it is necessary to rotate the standard frame of a
Skyrmion such that the vacuum is compatible with the potential
\eqref{eq:V} used here.
That is, a normal Skyrmion is a map $\bphi:\mathbb{R}^3\to S^3$ with
the vacuum chosen to be
\beq
U_{\rm vac}^{\textrm{standard frame}} = \mathbf{1}_2,
\eeq
which translates to the vacuum in $\bphi$:
\beq
\bphi_{\rm vac}^{\textrm{standard frame}} =
\begin{pmatrix}
  1\\
  0
\end{pmatrix},
\label{eq:vac_standard_frame}
\eeq
which is obviously not the vacuum \eqref{eq:vacuum} in this model,
i.e.~with the potential \eqref{eq:V}, but is indeed compatible with a
normal pion mass term $\tr(\mathbf{1}_2-U)$ or any generalization
thereof.  
Therefore, for all initial conditions, which are constructed to match
the standard frame, we perform the following rotation
\beq
\tilde\bphi = \frac{1}{\sqrt{2}}
\begin{pmatrix}
  1 & 1\\
  -1 & 1
\end{pmatrix} \Re(\bphi) + \Im(\bphi),
\label{eq:transformation}
\eeq
which is an SO(4) transformation transforming the standard frame
vacuum \eqref{eq:vac_standard_frame} into
\beq
\tilde\bphi_{\rm vac} = \frac{1}{\sqrt{2}}
\begin{pmatrix}
  1\\
  -1
\end{pmatrix}.
\eeq
This corresponds to the vacuum of this model in Eq.~\eqref{eq:vacuum}
with $\alpha=0$ and $\beta=\pi$. 

The initial conditions are now ready to be studied using the numerical
method described in the previous subsection and the results will be
given next.

\subsection{Numerical results}

We are now ready to present the numerical results in the 2+4 and
2+6 models for various values of $M$, with $M^2$ the coefficient of
the miscible BEC-inspired potential.

\def\scalefactor{0.24}%0.12%0.135%0.15
\def\scalefactorsix{0.16}%0.08%0.09%0.1
\begin{figure}[!htp]
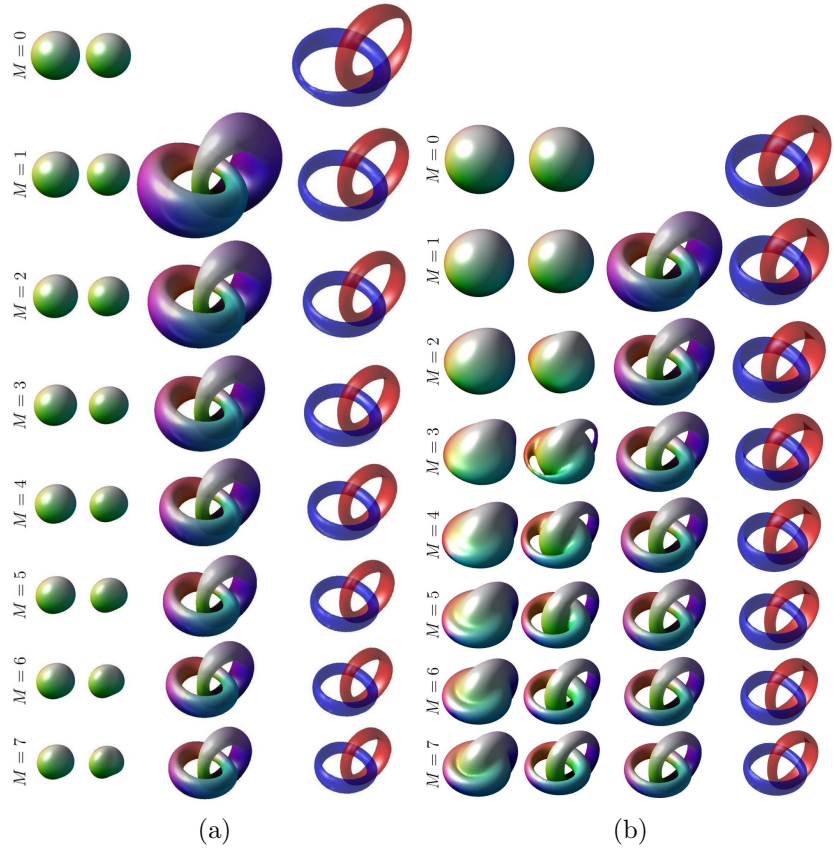

\begin{center}
  \mbox{\subfloat[]{\includegraphics[scale=\scalefactor]{{{mon1_0_small}}}}}
  \mbox{\subfloat[]{\includegraphics[scale=\scalefactorsix]{{{mon1_1_small}}}}}
  \caption{The $B=1$ Skyrmion in the miscible BEC-Skyrme (a) 2+4 model
    (b) 2+6 model.
    The four columns show isosurfaces of: the topological baryon
    charge density $\mathcal{B}$, the energy density $\mathcal{E}$,
    the potential $V$, and the vorticities $\mathcal{Q}_{1,2}$ in
    $\phi_{1,2}$ (with red and blue, respectively).
    The color scheme used for the first 3 columns is described in the
    text. Each row corresponds to a different value of the potential
    parameter $M^2$.
  }
  \label{fig:B1}
\end{center}
\end{figure}

Starting with the topological sector $B=1$, the soliton solution
called the Skyrmion, is spherically symmetric without the potential
(i.e.~for $M=0$).
The spherical symmetry means that a spatial SO(3) rotation can be
undone by an SU(2) isospin (internal) rotation.
Once $M\neq 0$ is turned on, the spherical symmetry is explicitly
broken.
The numerical results for $B=1$ with $M=0,1,\ldots,7$ are shown
Fig.~\ref{fig:B1} for both (a) the 2+4 model and (b) the 2+6 model. 
This figure and all the remaining figures of the same type in this
paper are composed of 4 columns showing isosurfaces of the topological
baryon charge density, of the total energy density, of the potential
and of the vorticities defined in Eq.~\eqref{eq:Q12}, respectively.
The color scheme utilized in the first 3 columns is the one used for
standard Skyrmions, where the three real components $\Im\phi_1$,
$\Re\phi_2$ and $\Im\phi_2$ are normalized
\beq
\begin{pmatrix}
  n_1\\
  n_2\\
  n_3
\end{pmatrix}
\equiv \frac{1}{(\Im\phi_1)^2 + |\phi_2|^2}
\begin{pmatrix}
  \Im\phi_1\\
  \Re\phi_2\\
  \Im\phi_2
\end{pmatrix},
\eeq
where $n_1+\i n_2=\exp(\i H)$ is mapped to the hue $H$ and $n_3$
determines the lightness, such that $n_3=1$ is white, $n_3=-1$ is
black and $n_3=0$ is a color determined by $H$: $H=0$ is red,
$H=2\pi/3$ is green and $H=4\pi/3$ is blue.
The intent is simply to show how the surfaces are mapped to the target
space $S^3$.
The last column in each panel of Fig.~\ref{fig:B1} shows the
vorticities given in Eq.~\eqref{eq:Q12} such that the vorticity
density in $\phi_1$ ($\phi_2$) is given by $Q_1$ ($Q_2$) and shown
with red (blue) surfaces.
All the isosurfaces are shown at half-maximum of the corresponding
observable, except for the vorticities, which are shown at a quarter
of the maximum vorticity. 

Fig.~\ref{fig:B1}(a) shows the $B=1$ Skyrmion for various values of
$M$ in the 2+4 model. It is interesting to note that the vorticities
in the 4th column are present with the potential turned off
($M=0$).\footnote{The 
  standard frame would make one of the vortices a ``vacuum'' vortex
  and the other a ``physical'' vortex, see
  Ref.~\cite{Gudnason:2020luj}. 
  The transformation \eqref{eq:transformation} is thus equivalent to
  one of the rotations of the 2-sphere performed in
  Ref.~\cite{Gudnason:2020luj}.}
Although both the potential (3rd column) and the vorticities clearly
show the 1-Skyrmion contains a pair of linked vortices for $M>0$, and
hence with linking number 1 -- in accord with the theorem of
Ref.~\cite{Gudnason:2020luj}, the linked vortices do not quite show
themselves in the topological baryon charge density or the total
energy density for the 2+4 model (Fig.~\ref{fig:B1}(a)).
All that happens is that the baryon charge density and energy density
isosurfaces are slightly deformed, which is expected due to the broken 
SU(2) symmetry.

Fig.~\ref{fig:B1}(b) shows the $B=1$ Skyrmion for various values in
the 2+6 model.
In comparison with the 2+4 model, the total energy density
isosurface (at half-maximum density) is quickly transformed from a
sphere into the shape given by two linked vortices, see the 2nd column
of Fig.~\ref{fig:B1}(b) -- this happens around $M\lesssim 3$. 
To some extent the linked vortices also become visible in the baryon
charge density, see the first column of the figure.

\begin{figure}[!htp]
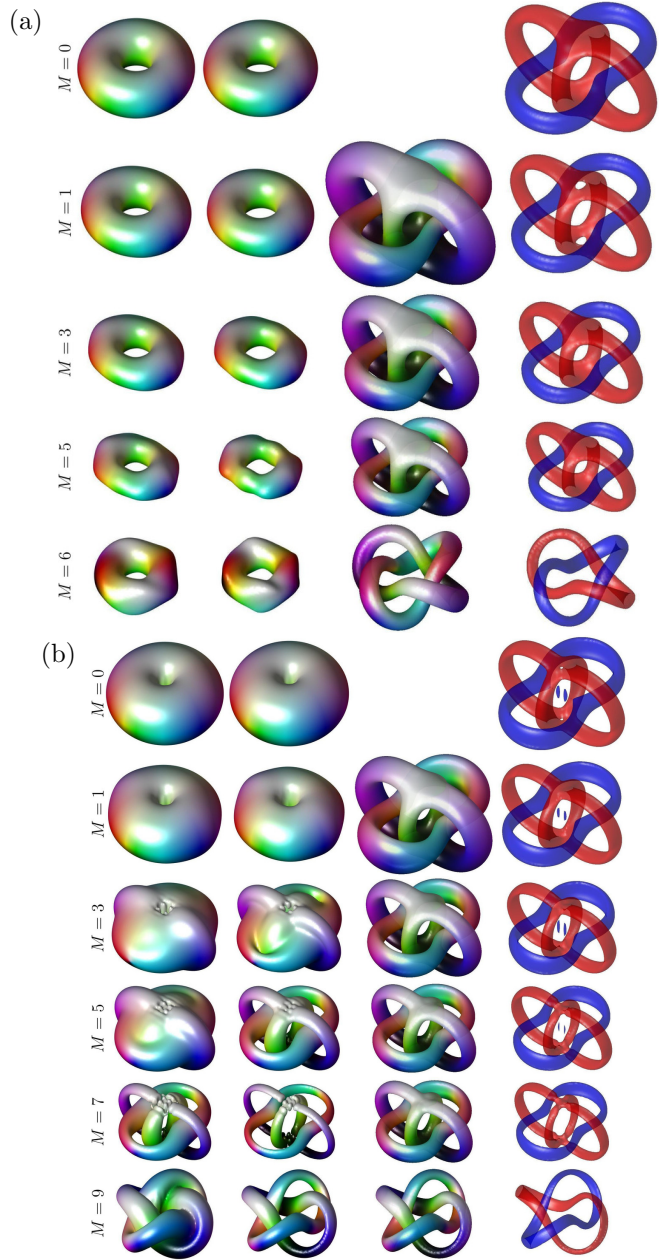

\begin{center}
  \mbox{\sidesubfloat[]{\includegraphics[scale=\scalefactor]{{{mon2_0_small}}}}}
  \mbox{\sidesubfloat[]{\includegraphics[scale=\scalefactorsix]{{{mon2_1_small}}}}}
  \caption{The \emph{metastable} $B=2$ Skyrmion in the miscible
    BEC-Skyrme (a) 2+4 model (b) 2+6 model.
    The four columns show isosurfaces of: the topological baryon
    charge density $\mathcal{B}$, the energy density $\mathcal{E}$,
    the potential $V$, and the vorticities $\mathcal{Q}_{1,2}$ in
    $\phi_{1,2}$ (with red and blue, respectively).
    The color scheme used for the first 3 columns is described in the
    text. Each row corresponds to a different value of the potential
    parameter $M^2$.
  }
  \label{fig:B2}
\end{center}
\end{figure}

We now turn to the case of the $B=2$ Skyrmion, which without a
potential is torus shaped.
Figs.~\ref{fig:B2}(a) and \ref{fig:B2}(b) show the results for
various masses $M$ in the 2+4 model and 2+6 model, respectively.
More precisely, we start with the standard $B=2$ Skyrmion transformed
by Eq.~\eqref{eq:transformation} as the initial guess.
We notice that the vorticities (4th column) are degenerate even for
$M=0$ and so is the potential energy (3rd column) once turning on a
finite potential, $M=1$.
In order to define what we mean by degenerate, let us make the
following definition:
\begin{definition}\label{def:1}
  Vortex links are degenerate, if they possess a mathematical junction
  which makes the counting of the linking number impossible.
\end{definition}
For a detailed discussion of the linking number, see
Ref.~\cite{Gudnason:2020luj}.
In particular, the vortex links are not necessarily degenerate, if at
a given level set, the isosurfaces merely touch each other.
That is, if the vortices do not touch each other by increasing the
level-set value from $1/2$ to a higher value of the maximum density,
then they are not degenerate by definition \ref{def:1}.

According to definition \ref{def:1}, the $B=2$ Skyrmion shown in
Fig.~\ref{fig:B2} is \emph{degenerate} until a sufficiently high value
of the potential parameter is reached: i.e.~at $M\sim 5$ for the 2+4
model and $M\sim 9$ for the 2+6 model the degeneracy is broken
spontaneously and the vortex links fall into a nondegenerate state
where the linking number is clearly two -- this can be seen from the
potential (3rd column) and the vorticities (4th column) of the
figure. 
For the 2+4 model, the baryon charge and the total energy densities
are only slightly deformed, whereas for the 2+6 model once the
degeneracy is broken, the baryon charge shows clear signs of the linked
vortices inside the Skyrmion and the energy density takes the same
shape as the potential -- i.e.~as two doubly linked vortices.

\begin{figure}[!htp]
\begin{center}
  \mbox{\sidesubfloat[]{\includegraphics[scale=\scalefactor]{{{mon2_0_B2_0down_small}}}}}
  \mbox{\sidesubfloat[]{\includegraphics[scale=\scalefactorsix]{{{mon2_1_B2_1down_small}}}}}
  \caption{The \emph{stable} $B=2$ Skyrmion in the miscible
    BEC-Skyrme (a) 2+4 model (b) 2+6 model.
    The four columns show isosurfaces of: the topological baryon
    charge density $\mathcal{B}$, the energy density $\mathcal{E}$,
    the potential $V$, and the vorticities $\mathcal{Q}_{1,2}$ in
    $\phi_{1,2}$ (with red and blue, respectively).
    The color scheme used for the first 3 columns is described in the
    text. Each row corresponds to a different value of the potential
    parameter $M^2$.
  }
  \label{fig:B2stable}
\end{center}
\end{figure}

\begin{figure}[!htp]
  \begin{center}
    \includegraphics[width=0.5\linewidth]{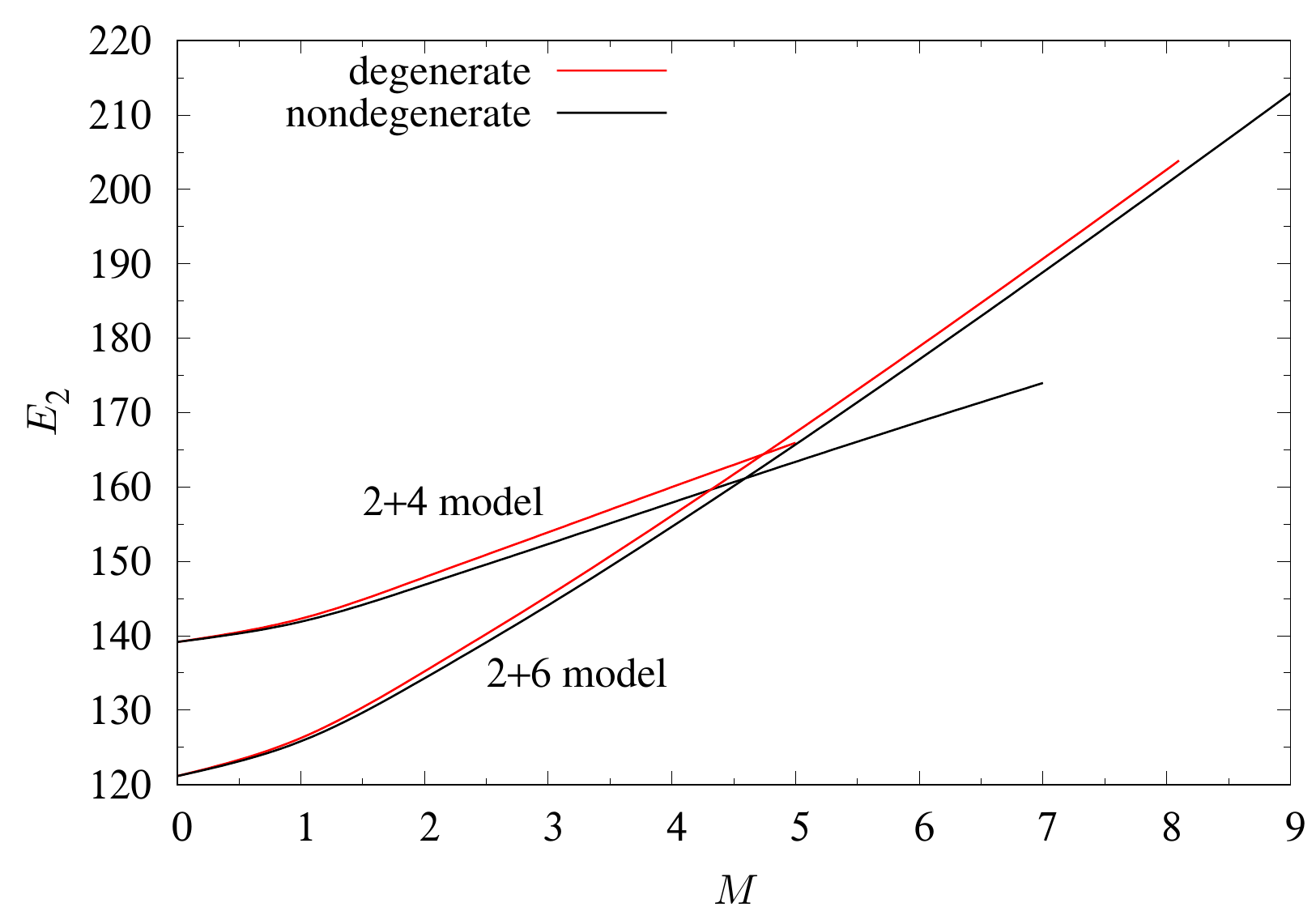}
    \caption{Energy of degenerate versus nondegenerate $B=2$ Skyrmion
      solutions for both the 2+4 model and the 2+6 model. } 
    \label{fig:en2}
  \end{center}
\end{figure}

One could now expect that the degenerate soliton becomes unstable for
a sufficiently large $M$ and decays into the nondegenerate state.
This begs the question, what is the phase diagram of these two states
as function of $M$.
In particular, which is the stable state for smaller values of $M$.
In order to investigate this, we will use the $B=2$ Skyrmion with
broken degeneracy (see the last lines of Figs.~\ref{fig:B2}(a,b)) as
the initial guess and decrease $M$ to zero -- and while doing so, we
calculate the total energy.
The result is shown in Fig.~\ref{fig:B2stable} and for both models,
the linked vortices are nondegenerate all the way as $M$ tends to
zero, thus allowing one to count the topological degree as the linking
number of the two vortex species, see Ref.~\cite{Gudnason:2020luj}.
In order to conclude which of the two solutions is the stable one, we
plot the energies in Fig.~\ref{fig:en2}.
Indeed, the solutions with nondegenerate vortices turn out to be the
stable ones.
For the stable solutions, the linked vortices are visible in the
energy density from $M\gtrsim 3$ in the 2+6 model, whereas for the
2+4 model they are only barely visible at $M=6$. 

One could posit that starting with the (transformed) rational map
approximated Skyrmion is the reason for ending up in the metastable
state.
Hence, we have performed a scattering of two $B=1$ Skyrmions in the
attractive channel.
The initial state was prepared by means of the asymmetric product
Ansatz of two 1-Skyrmions, which was then transformed into the frame
compatible with this model's vacuum using
Eq.~\eqref{eq:transformation}.
Surprisingly, the result for $M=1$ was the metastable
``degenerate'' state instead of the stable ground state.

\begin{figure}[!htp]
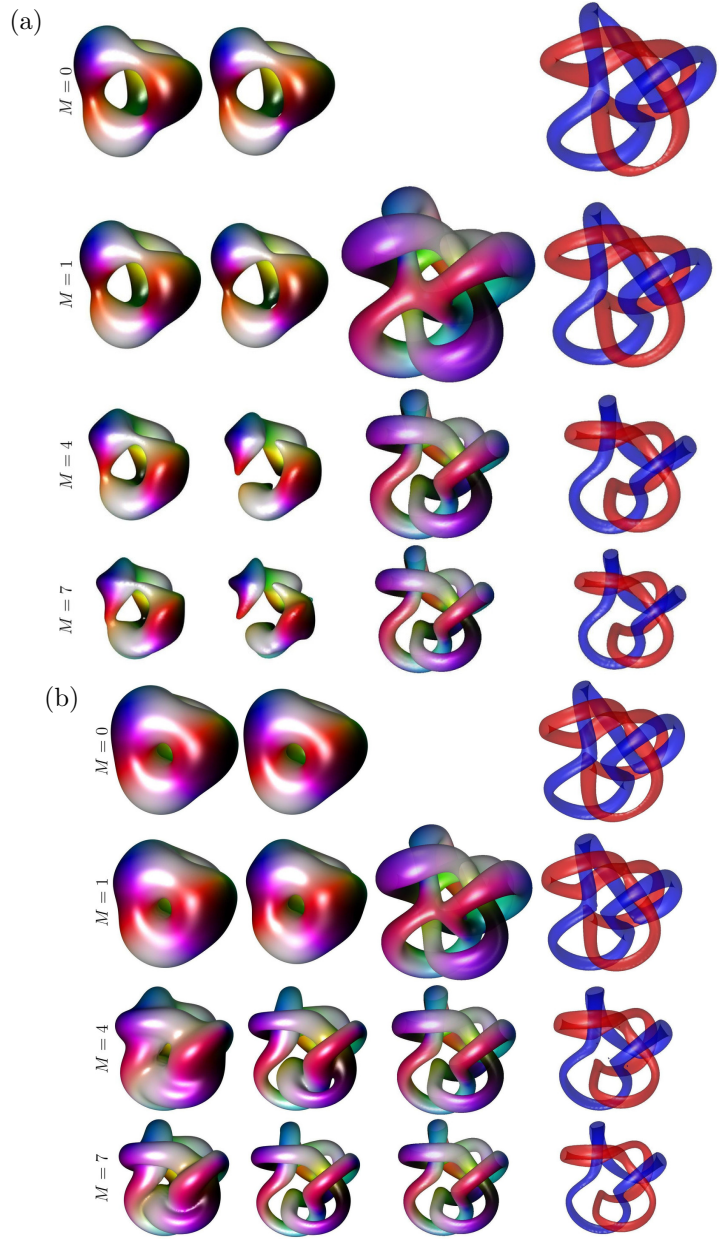

\begin{center}
  \mbox{\sidesubfloat[]{\includegraphics[scale=\scalefactor]{{{mon3_0_small}}}}}
  \mbox{\sidesubfloat[]{\includegraphics[scale=\scalefactorsix]{{{mon3_1_small}}}}}
  \caption{The $B=3$ Skyrmion in the miscible
    BEC-Skyrme (a) 2+4 model (b) 2+6 model.
    The four columns show isosurfaces of: the topological baryon
    charge density $\mathcal{B}$, the energy density $\mathcal{E}$,
    the potential $V$, and the vorticities $\mathcal{Q}_{1,2}$ in
    $\phi_{1,2}$ (with red and blue, respectively).
    The color scheme used for the first 3 columns is described in the
    text. Each row corresponds to a different value of the potential
    parameter $M^2$.
  }
  \label{fig:B3}
\end{center}
\end{figure}

We will now consider the $B=3$ topological sector, for which the
results are shown in Fig.~\ref{fig:B3}.
The $M=0$ row of Fig.~\ref{fig:B3}(a) (Fig.~\ref{fig:B3}(b)) shows the
normal tetrahedrally symmetric Skyrmion in the 2+4 (2+6) model.
The triply linked vortex lines are visible in the 4th column of the
figure and they are not degenerate for this soliton solution.
Turning on a finite $M$ yields the potential energy (3rd column) in a
similar shape as the vorticities (4th column).
For $M=1$ it may look like the potential (3rd column) shows signs of
degeneracy, but this is merely an illusion due to the level set and is
not a degeneracy according to definition \ref{def:1}; this can be seen
by inspecting the vorticities in the fourth column of the figure. 
In the 2+4 model, the energy density isosurface changes for large
values of $M$ (i.e.~$M\sim4$--$7$), but instead of becoming similar to
the shape of the potential isosurface (3rd column) it separates into
four lumps centered at the corners of the tetrahedron with hints of
the vortex lines sticking out.
In contrast, for the 2+6 model, the energy density isosurface becomes
very similar in shape to that of the potential energy for large values
of $M$ (i.e.~$M\sim4$--$7$), see the second column of
Fig.~\ref{fig:B3}.
This can be traced back to the perfect fluid properties of the
BPS-Skyrme term \cite{Adam:2014nba}.

\begin{figure}[!htp]
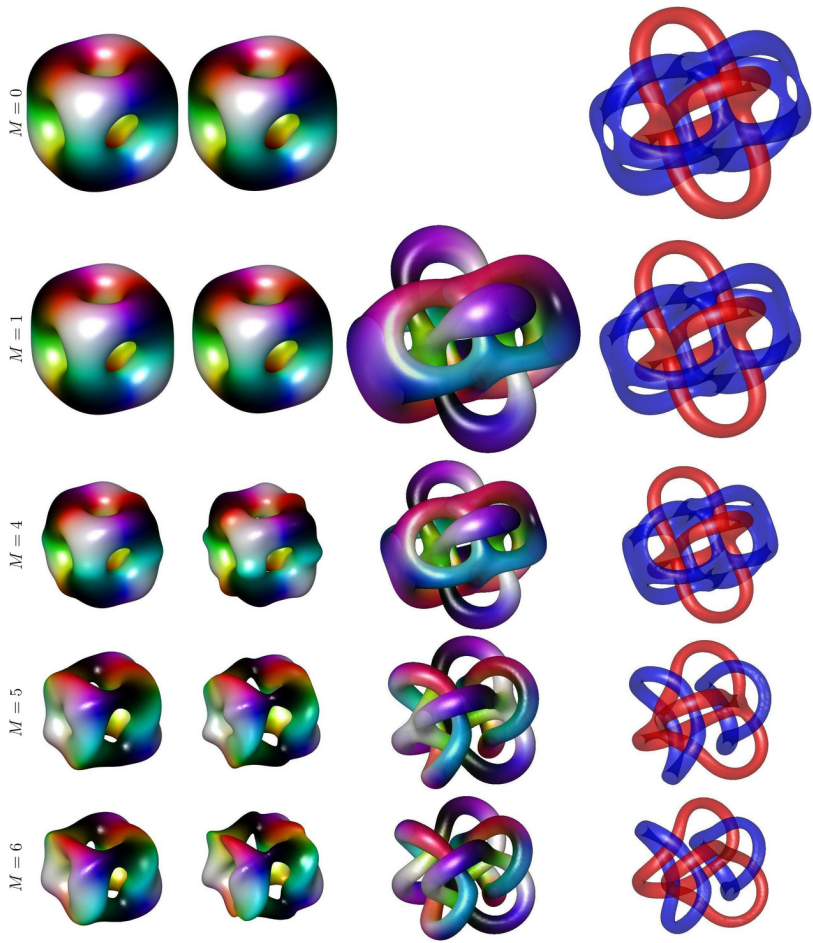

\begin{center}
  \includegraphics[scale=\scalefactor]{{{mon4_0_small}}}
  \caption{The \emph{metastable} $B=4$ Skyrmion in the miscible
    BEC-Skyrme 2+4 model. 
    The four columns show isosurfaces of: the topological baryon
    charge density $\mathcal{B}$, the energy density $\mathcal{E}$,
    the potential $V$, and the vorticities $\mathcal{Q}_{1,2}$ in
    $\phi_{1,2}$ (with red and blue, respectively).
    The color scheme used for the first 3 columns is described in the
    text. Each row corresponds to a different value of the potential
    parameter $M^2$.
  }
  \label{fig:B4a}
  \end{center}
\end{figure}

\begin{figure}[!htp]
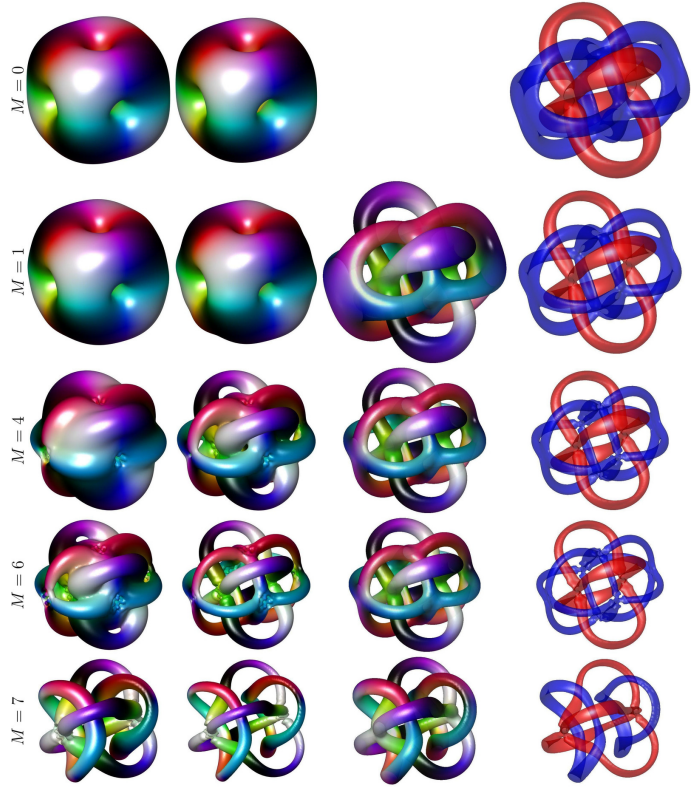

\begin{center}
  \includegraphics[scale=\scalefactorsix]{{{mon4_1_small}}}
  \caption{The \emph{metastable} $B=4$ Skyrmion in the miscible
    BEC-Skyrme 2+6 model.
    The four columns show isosurfaces of: the topological baryon
    charge density $\mathcal{B}$, the energy density $\mathcal{E}$,
    the potential $V$, and the vorticities $\mathcal{Q}_{1,2}$ in
    $\phi_{1,2}$ (with red and blue, respectively).
    The color scheme used for the first 3 columns is described in the
    text. Each row corresponds to a different value of the potential
    parameter $M^2$.
  }
  \label{fig:B4b}
\end{center}
\end{figure}

We now turn to the $B=4$ sector, where the Skyrmion solution for $M=0$
is octahedrally symmetric (which is the dual symmetry of the cube).
First we obtain solutions for various values of $M$ using the initial
data which are transformed rational map solutions for the 2+4 model in
Fig.~\ref{fig:B4a} and for the 2+6 model in Fig.~\ref{fig:B4b}.
In both the 2+4 and the 2+6 model, the solution obtained for small
values of $M$ is degenerate according to definition \ref{def:1}, see
the 3rd and 4th columns of Figs.~\ref{fig:B4a} and \ref{fig:B4b},
respectively.
The critical value of the mass parameter is $M_{\rm crit}\sim 4.3$ for
the 2+4 model and $M_{\rm crit}\sim 6$ for the 2+6 model.
Interestingly, the breaking of degeneracy that happens at this value
of $M$ is only partial.
That is, the degeneracy of the blue vortices breaks spontaneously, but
that of the red vortices sustains, see the fourth column in
Figs.~\ref{fig:B4a} and \ref{fig:B4b}. 
The lower-in-energy metastable state, with only the red vortices being
degenerate, can be seen for $M=5$ in Fig.~\ref{fig:B4a} and for $M=7$
in Fig.~\ref{fig:B4b}.
If we continue to increase the mass $M$, we expect to be able to break
the lower-in-energy metastable state with only the red vortices being
degenerate, and indeed that happens at $M_{\rm crit}'\sim 5.4$ in the
2+4 model. For the 2+6 model, we have not been able to find this
critical value of the mass parameter; by increasing $M$ to $M=15$, the
lower-in-energy metastable state is still metastable.
Beyond that value of the mass parameter, we do not trust the accuracy
of the code for the lattices used in these simulations.

\begin{figure}[!htp]
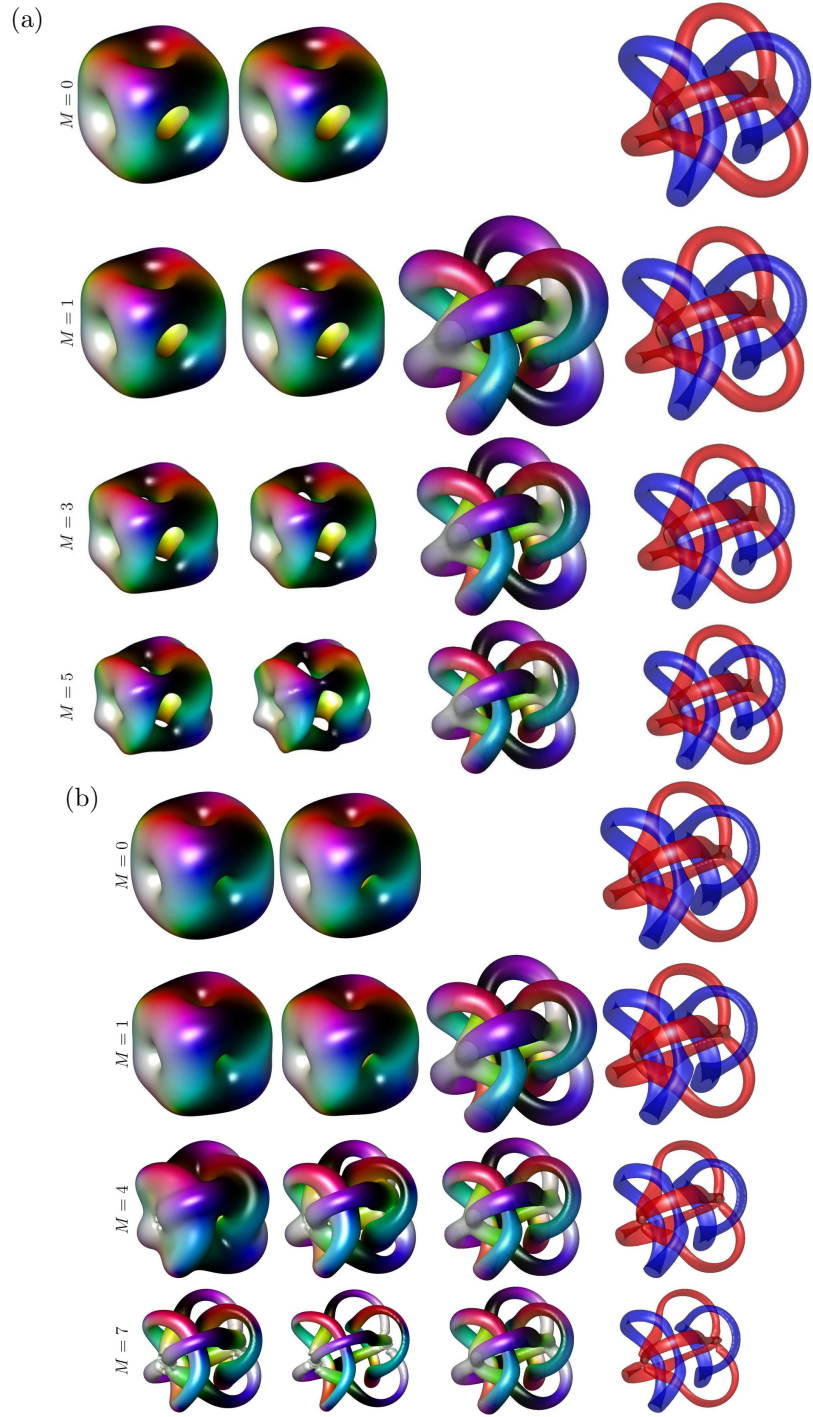

\begin{center}
  \mbox{\sidesubfloat[]{\includegraphics[scale=\scalefactor]{{{mon4_0_B4_0down2_small}}}}}
  \mbox{\sidesubfloat[]{\includegraphics[scale=\scalefactorsix]{{{mon4_1_B4_1down2_small}}}}}
  \caption{The lower-in-energy \emph{metastable} $B=4$ Skyrmion in
    the miscible BEC-Skyrme (a) 2+4 model (b) 2+6 model.
    The four columns show isosurfaces of: the topological baryon
    charge density $\mathcal{B}$, the energy density $\mathcal{E}$,
    the potential $V$, and the vorticities $\mathcal{Q}_{1,2}$ in
    $\phi_{1,2}$ (with red and blue, respectively).
    The color scheme used for the first 3 columns is described in the
    text. Each row corresponds to a different value of the potential
    parameter $M^2$.
  }
  \label{fig:B4metastable}
\end{center}
\end{figure}

\begin{figure}[!htp]
\begin{center}
  \mbox{\sidesubfloat[]{\includegraphics[scale=\scalefactor]{{{mon4_0_B4_0down_small}}}}}
  \mbox{\sidesubfloat[]{\includegraphics[scale=\scalefactorsix]{{{mon4_1_B4_1down_small}}}}}
  \caption{The \emph{stable} $B=4$ Skyrmion in
    the miscible BEC-Skyrme (a) 2+4 model (b) 2+6 model.
    The four columns show isosurfaces of: the topological baryon
    charge density $\mathcal{B}$, the energy density $\mathcal{E}$,
    the potential $V$, and the vorticities $\mathcal{Q}_{1,2}$ in
    $\phi_{1,2}$ (with red and blue, respectively).
    The color scheme used for the first 3 columns is described in the
    text. Each row corresponds to a different value of the potential
    parameter $M^2$.
  }
  \label{fig:B4stable}
\end{center}
\end{figure}

\begin{figure}[!htp]
  \begin{center}
    \mbox{\subfloat[]{\includegraphics[width=0.5\linewidth]{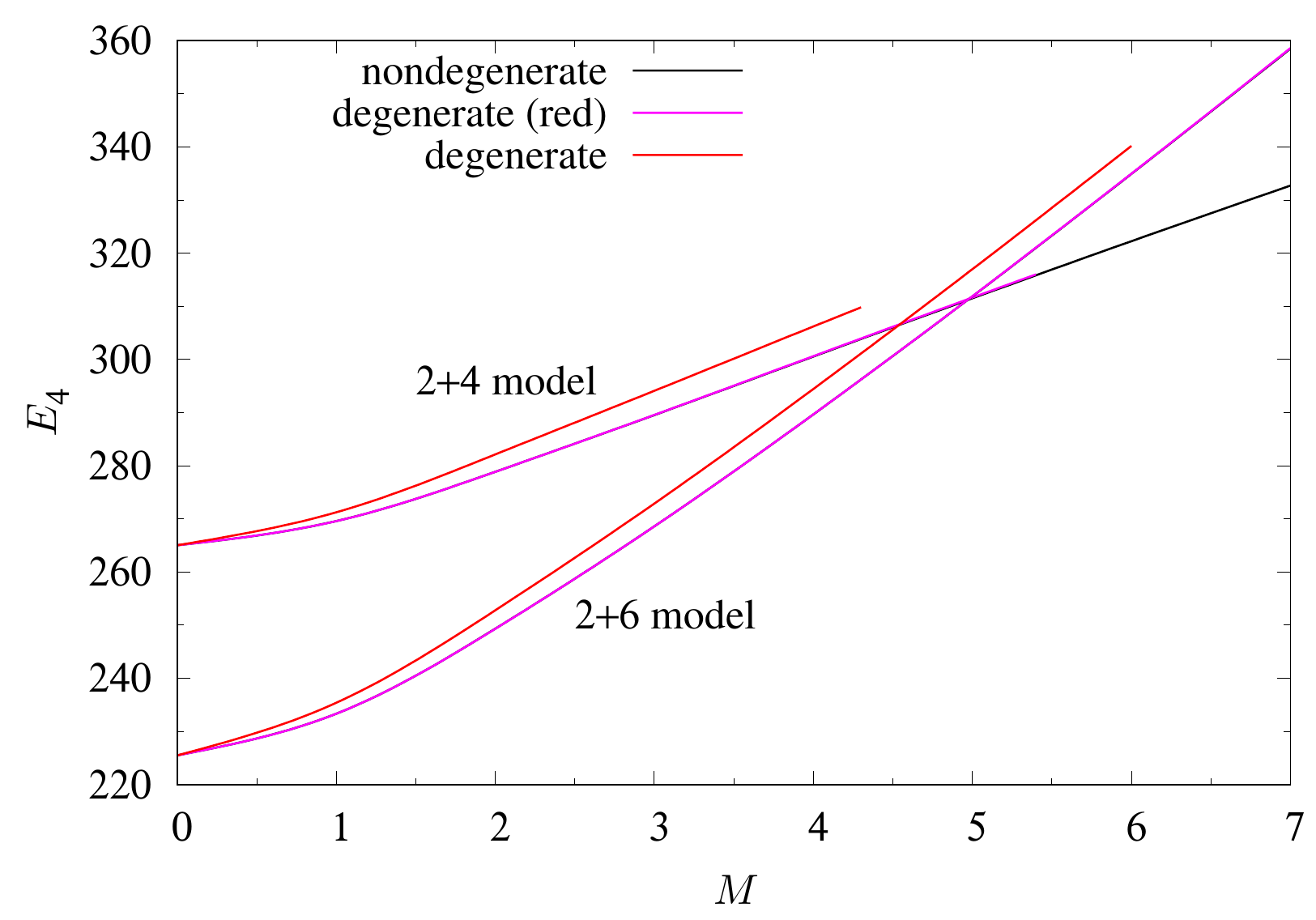}}
      \subfloat[]{\includegraphics[width=0.5\linewidth]{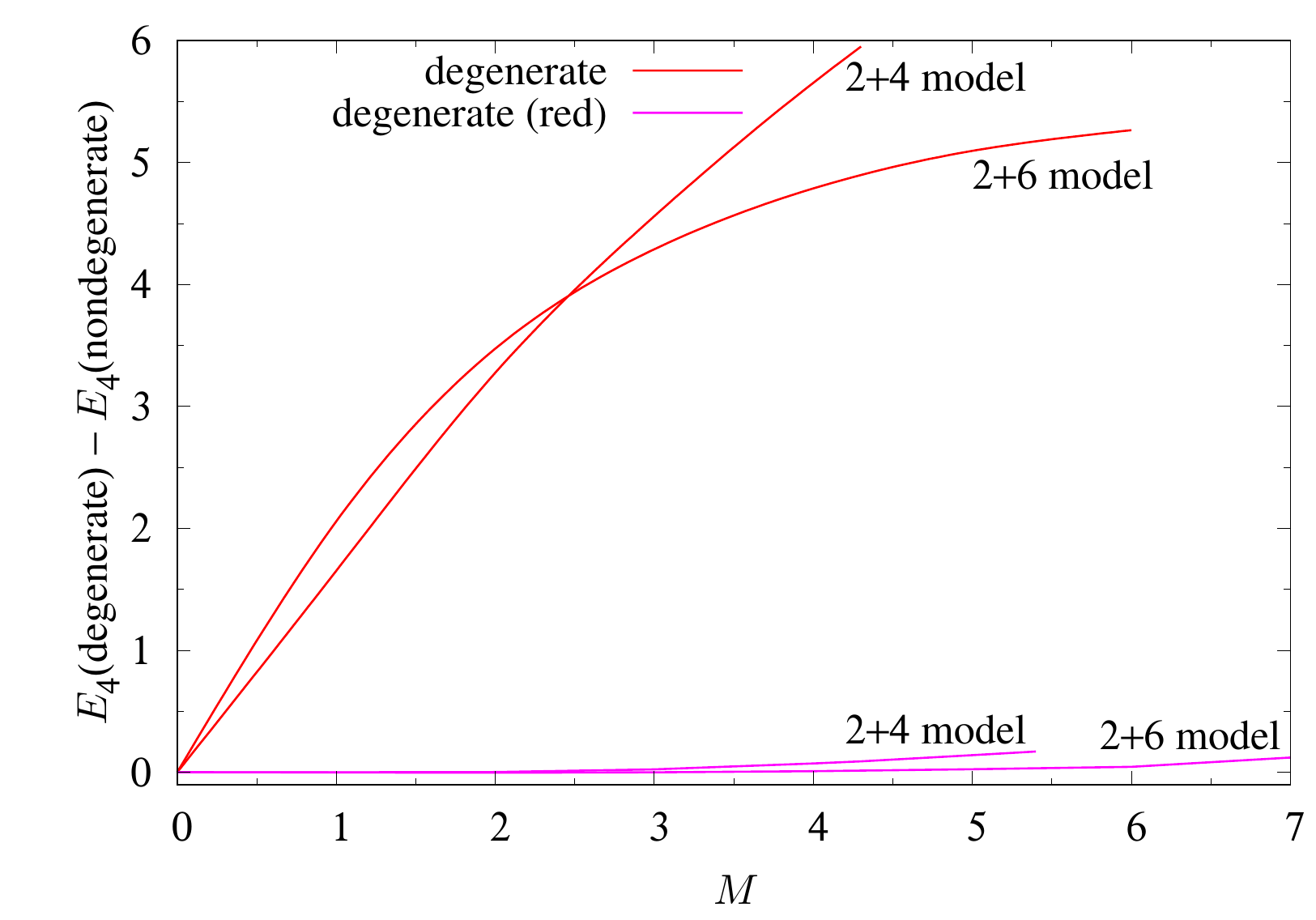}}}
    \caption{(a) Energy of degenerate versus nondegenerate $B=4$ Skyrmion
      solutions for both the 2+4 model and the 2+6 model.
      There is a degenerate state with almost the same energy as the
      nondegenerate state, where only the red vortex $(\phi_1)$ is
      degenerate, but the blue vortex $(\phi_2)$ is nondegenerate.
      (b) The energy difference between the degenerate states and the
      nondegenerate ``ground state''. 
    } 
    \label{fig:en4}
  \end{center}
\end{figure}

Fig.~\ref{fig:B4metastable} shows the lower-in-energy metastable
states, where only the red vortices are degenerate while the blue
vortices are nondegenerate, see the fourth column of the figure. 
Fig.~\ref{fig:B4stable} shows the stable nondegenerate soliton
solution, which is nondegenerate in both the red $(\phi_1)$ and blue
$(\phi_2)$ vortices.
In order to back up our claim, we calculate the energies for the
different metastable and the stable solutions as functions of $M$ and
the result is shown in Fig.~\ref{fig:en4}.
It is clear both from the energies in Fig.~\ref{fig:en4} and from
inspecting the baryon charge (1st column) and energy (2nd column)
isosurfaces of Figs.~\ref{fig:B4a}-\ref{fig:B4stable}, that the $M=0$
solution is energetically and physically the same solution.
Nevertheless, the vorticities plots are clearly different in
Figs.~\ref{fig:B4a}-\ref{fig:B4b} versus Fig.~\ref{fig:B4metastable}
versus Fig.~\ref{fig:B4stable}.
This is because, for vanishing potential, the difference is merely a
rotation of the 2-sphere in the language of
Ref.~\cite{Gudnason:2020luj}, which is a subgroup of O(4) and hence
does not change the energy or the physics of the Skyrmion with
massless pions.
Once, the potential \eqref{eq:V} is turned on, this rotation is no
longer a symmetry and the potential picks out the nondegenerate vortex
links as the ground state.
Perhaps surprisingly, there are metastable states with residual
degeneracy or even partial residual degeneracy left.
This was not expected.

Although the fully degenerate state shown in
Figs.~\ref{fig:B4a}-\ref{fig:B4b} is clearly heavier on the energy
graph \ref{fig:en4}(a) than the other two solutions, the difference
between the lower-in-energy metastable state of
Fig.~\ref{fig:B4metastable} and the stable nondegenerate solution of
Fig.~\ref{fig:B4stable} is actually so small, that we have plotted the
difference between the energies of the degenerate and the
nondegenerate states in Fig.~\ref{fig:en4}(b).
Clearly the partially degenerate solution with only the red vortices
being degenerate, has almost the same energy as the nondegenerate
stable solution.
This was not expected in the model with the potential \eqref{eq:V}
turned on.

We will now summarize the numerical results so far.
The $B=1$ case does not possess degeneracy in any frame, see
Ref.~\cite{Gudnason:2020luj}, but the second vortex is not necessarily
physical, as it can go off to infinity as a ``vacuum'' vortex.
For $B=2$ the standard frame after the transformation
\eqref{eq:transformation} turns out to give a 
degenerate solution. Cranking up the potential parameter turns this
metastable degenerate state into an unstable state and the true
nondegenerate ``ground state'' is found.
For $B=3$ the found solution does not possess degeneracy.
For $B=4$ the standard frame after rotation yields a metastable
state with degeneracy which is higher in energy than another
metastable state with partial degeneracy.
The stable solution, however, in all cases is the nondegenerate one.

\begin{figure}[!htp]
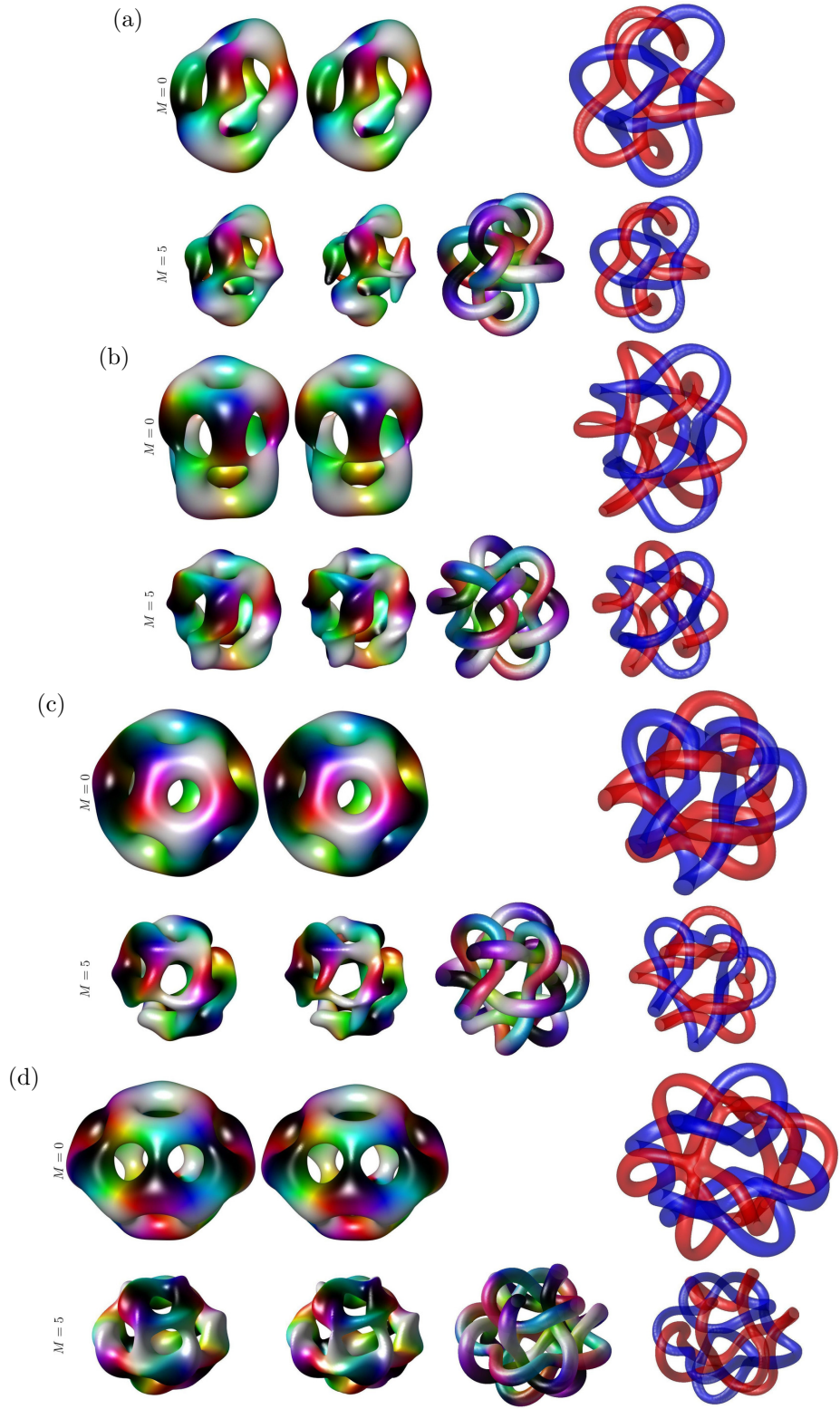

\begin{center}
  \renewcommand{\scalefactor}{0.204}%0.102
  \renewcommand{\scalefactorsix}{0.136}%0.068
  \mbox{\sidesubfloat[]{\includegraphics[scale=\scalefactor]{{{mon5_0_B5_0down_small}}}}}
  \mbox{\sidesubfloat[]{\includegraphics[scale=\scalefactor]{{{mon6_0_small}}}}}
  \mbox{\sidesubfloat[]{\includegraphics[scale=\scalefactor]{{{mon7_0_small}}}}}
  \mbox{\sidesubfloat[]{\includegraphics[scale=\scalefactor]{{{mon8_0_B8_0down_small}}}}}
  \caption{The \emph{stable} (a) $B=5$, (b) $B=6$, (c) $B=7$, (d)
    $B=8$ Skyrmions in the miscible BEC-Skyrme 2+4 model.
    %% The four columns show isosurfaces of: the topological baryon
    %% charge density $\mathcal{B}$, the energy density $\mathcal{E}$,
    %% the potential $V$, and the vorticities $\mathcal{Q}_{1,2}$ in
    %% $\phi_{1,2}$ (with red and blue, respectively).
    %% The color scheme used for the first 3 columns is described in the
    %% text. Each row corresponds to a different value of the potential
    %% parameter $M^2$.
  }
  \label{fig:B5-8_2+4}
  \end{center}
\end{figure}

\begin{figure}[!htp]
\begin{center}
  \mbox{\sidesubfloat[]{\includegraphics[scale=\scalefactorsix]{{{mon5_1_B5_1down_small}}}}}
  \mbox{\sidesubfloat[]{\includegraphics[scale=\scalefactorsix]{{{mon6_1_small}}}}}
  \mbox{\sidesubfloat[]{\includegraphics[scale=\scalefactorsix]{{{mon7_1_small}}}}}
  \mbox{\sidesubfloat[]{\includegraphics[scale=\scalefactorsix]{{{mon8_1_B8_1down_small}}}}}
  \caption{The \emph{stable} (a) $B=5$, (b) $B=6$, (c) $B=7$, (d)
    $B=8$ Skyrmions in the miscible BEC-Skyrme 2+6 model.
    The four columns show isosurfaces of: the topological baryon
    charge density $\mathcal{B}$, the energy density $\mathcal{E}$,
    the potential $V$, and the vorticities $\mathcal{Q}_{1,2}$ in
    $\phi_{1,2}$ (with red and blue, respectively).
    The color scheme used for the first 3 columns is described in the
    text. Each row corresponds to a different value of the potential
    parameter $M^2$.
  }
  \label{fig:B5-8_2+6}
\end{center}
\end{figure}

Since we have shown the first four Skyrmion solutions in great detail,
we will only depict the stable ones of the next four Skyrmions ($B=5$
through $B=8$) and only for $M=0,5$: For the 2+4 (2+6) model the
results are shown in Fig.~\ref{fig:B5-8_2+4}
(Fig.~\ref{fig:B5-8_2+6}).
The stable Skyrmions without the potential (i.e.~with $M=0$) have
dihedral $D_{2d}$, dihedral $D_{4d}$, icosahedral and dihedral
$D_{6d}$ symmetry, respectively.

\begin{figure}[!htp]
  \begin{center}
    \mbox{\subfloat[$B=5$]{\includegraphics[width=0.5\linewidth]{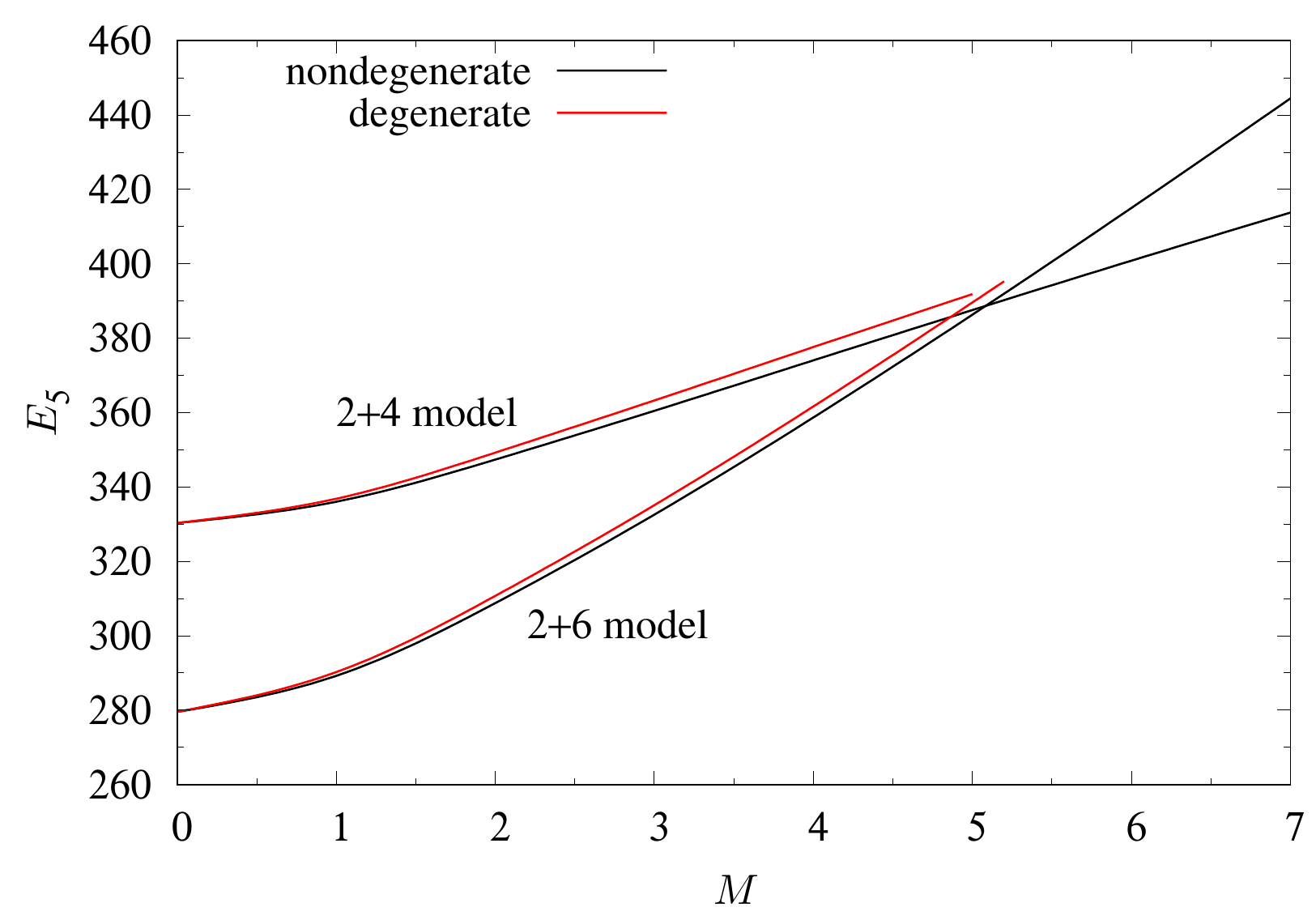}}
      \subfloat[$B=8$]{\includegraphics[width=0.5\linewidth]{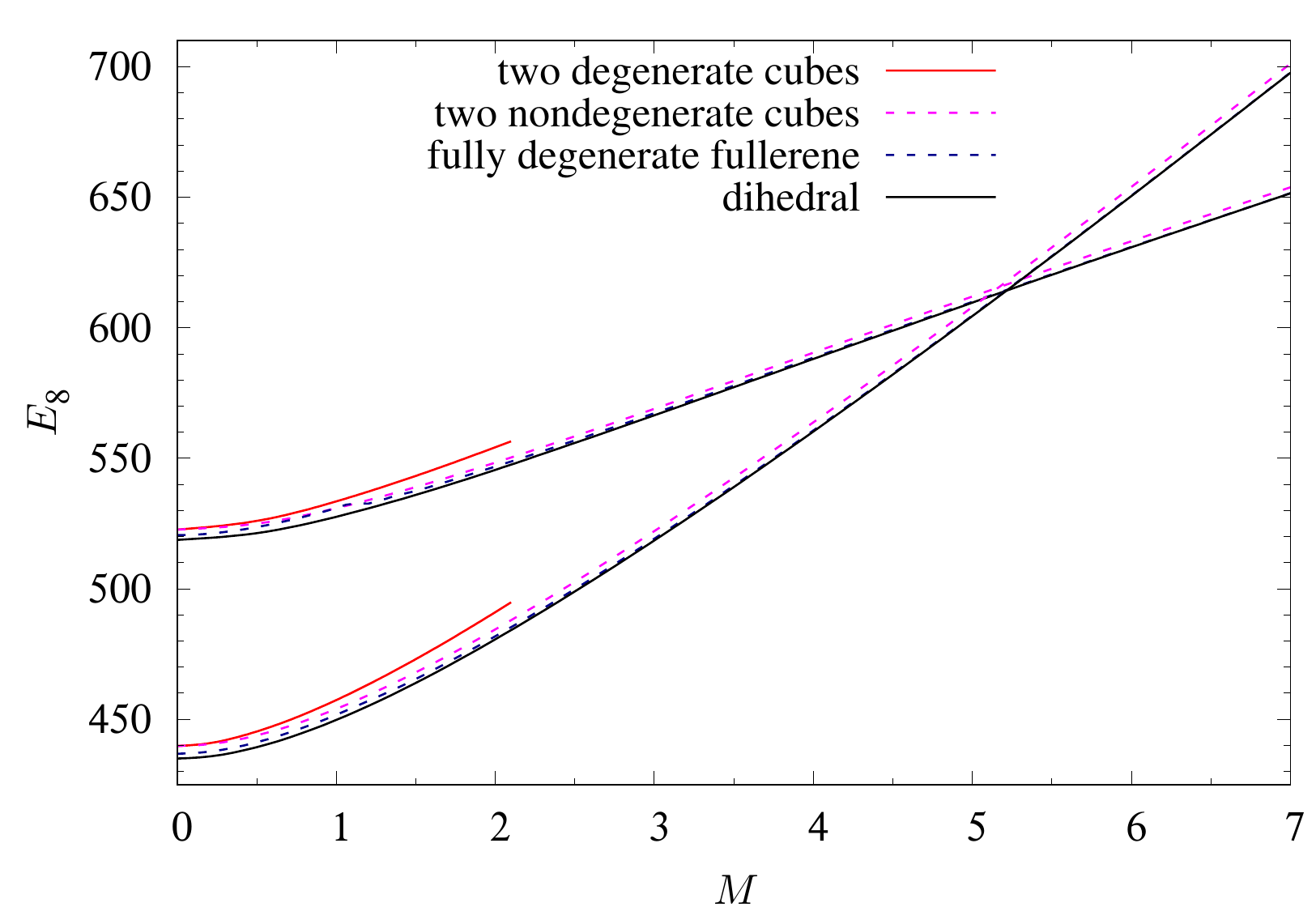}}}
    \mbox{\subfloat[$B=8$: 2+4 model]{\includegraphics[width=0.5\linewidth]{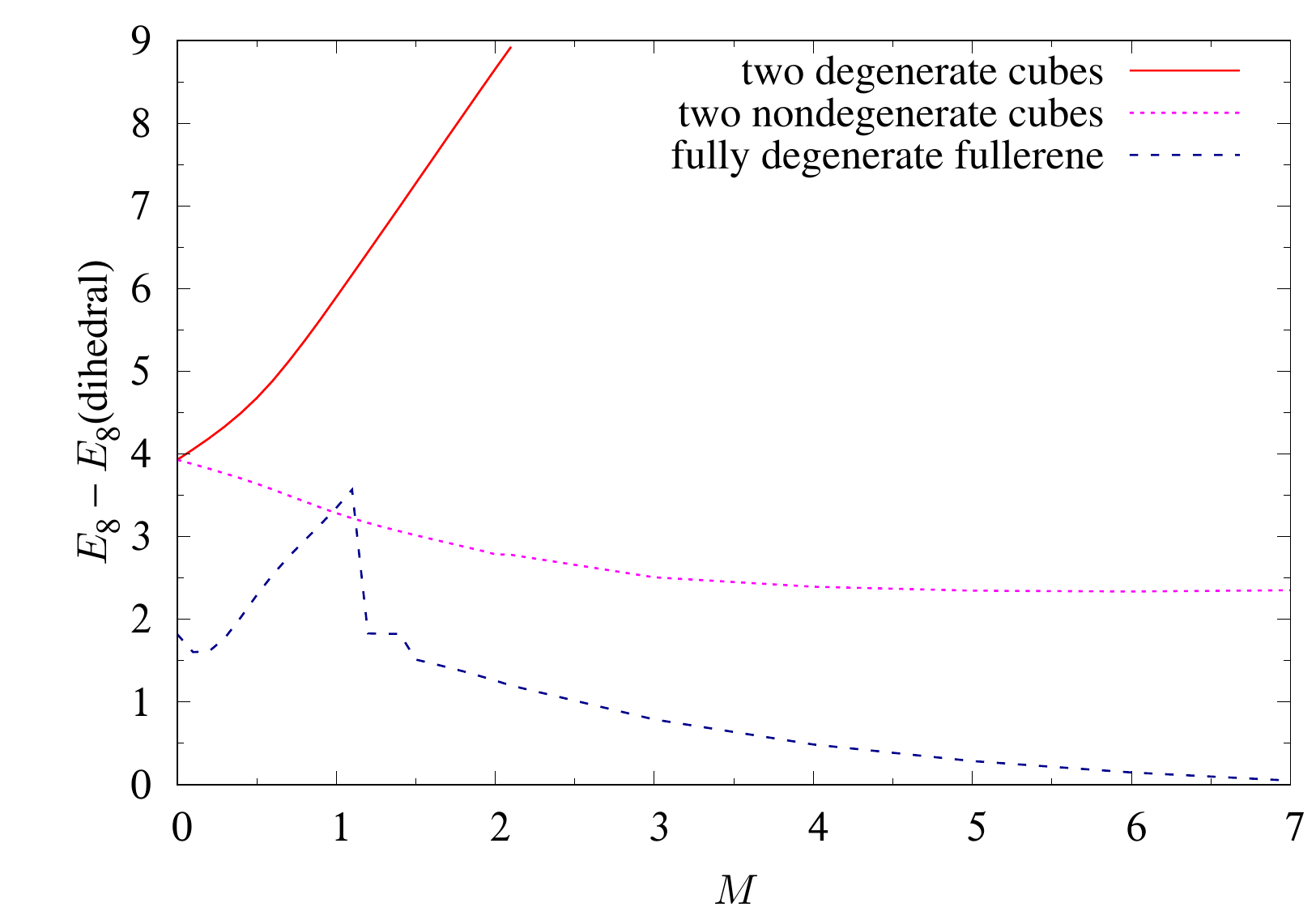}}
      \subfloat[$B=8$: 2+6 model]{\includegraphics[width=0.5\linewidth]{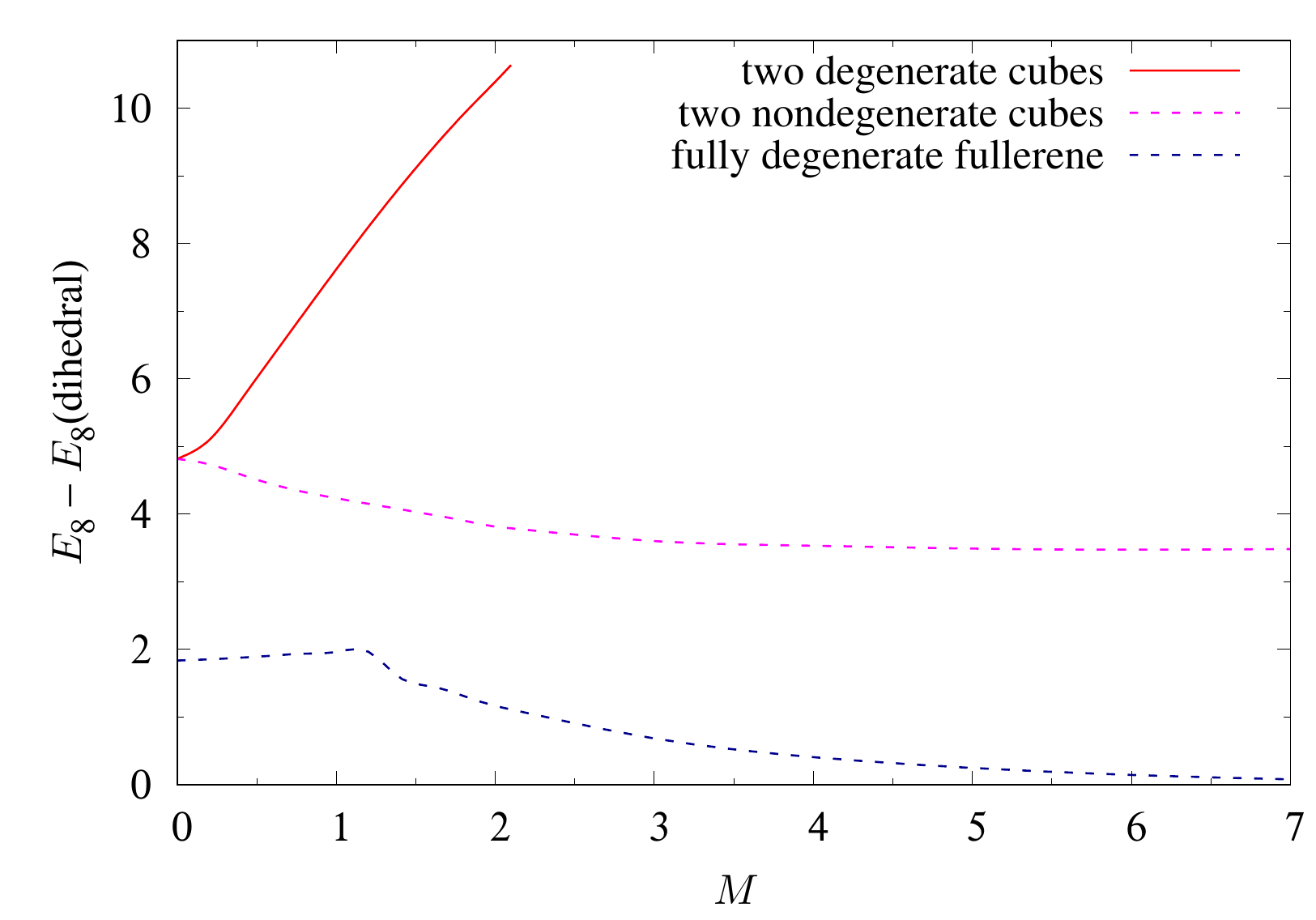}}}
    \caption{Energy of degenerate versus nondegenerate (a) $B=5$ and
      (b) $B=8$ Skyrmion solutions for both the 2+4 model and the 2+6
      model.
      A difference plot between the stable ``dihedral'' solution
      and the other metastable solutions for (c) the 2+4 model and (d)
      the 2+6 model. 
    } 
    \label{fig:en58}
  \end{center}
\end{figure}

It turns out that for $B=5$, there is a stable solution which
possesses nondegenerate vortex links, see Figs.~\ref{fig:B5-8_2+4}(a)
and \ref{fig:B5-8_2+6}(a), as well as an unstable solution with
degenerate links (not shown).
The energies of the two branches of solutions are shown in
Fig.~\ref{fig:en58}(a) and the situation is very similar to that of
the $B=2$ Skyrmion.

For the $B=6,7$ Skyrmions, we only find a stable solution shown in
Figs.~\ref{fig:B5-8_2+4}(b,c) and \ref{fig:B5-8_2+6}(b,c) for the 2+4
model and the 2+6 model, respectively.
Whereas the $B=7$ solution is clearly nondegenerate, the $B=6$
solution looks almost degenerate, see the fourth column of the figures
for $M=0$. This turns out to be an artifact of the level set and
indeed the degeneracy according to definition \ref{def:1} is not
present.

The last Skyrmion solution is the $B=8$ solution, for which the stable
Skyrmion in this model is shown in Figs.~\ref{fig:B5-8_2+4}(d) and
\ref{fig:B5-8_2+6}(d) for the 2+4 model and the 2+6 model,
respectively. 
For all the Skyrmion solutions in the 2+4 model, see
Fig.~\ref{fig:B5-8_2+4}, the vortex links are less visible than in
the 2+6 model, see Fig.~\ref{fig:B5-8_2+6}, even for large $M$.
That being said, the deformations of the Skyrmion solutions in the 2+4
model are definitely visible for large values of $M$, but in the 2+6
model, the shape of the energy density (2nd column) converges quickly
to that of the potential (3rd column), which by comparison to the
vorticities (4th column) represents the two flavors of vortex, which
when nondegenerate are linked exactly $B$ times
\cite{Gudnason:2020luj}.

\begin{figure}[!htp]
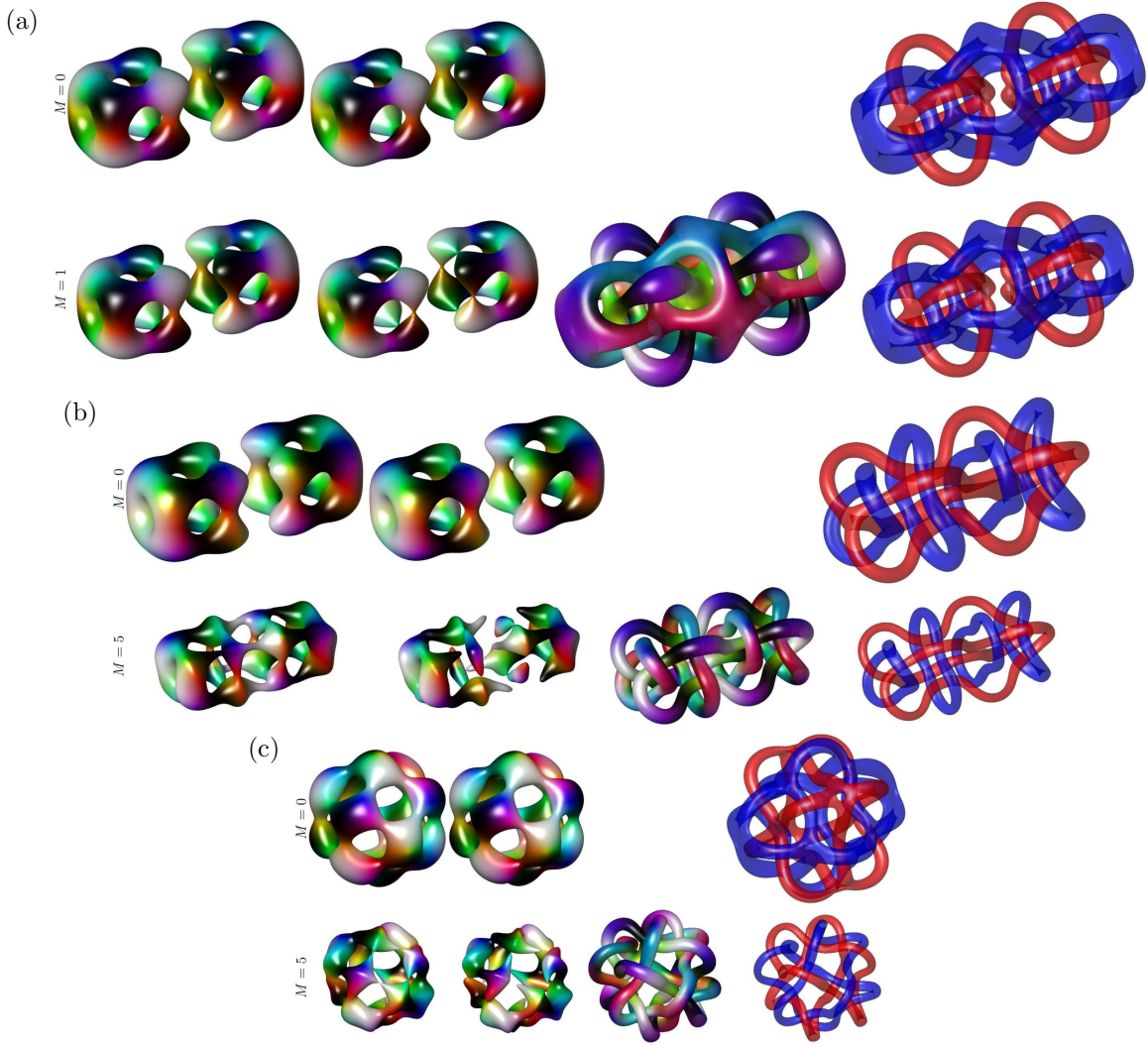

\begin{center}
  \renewcommand{\scalefactor}{0.204}%0.102
  \renewcommand{\scalefactorsix}{0.136}%0.068
  \mbox{\sidesubfloat[]{\includegraphics[scale=\scalefactor]{{{mon8_0_B8_0cubes_small}}}}}
  \mbox{\sidesubfloat[]{\includegraphics[scale=\scalefactor]{{{mon8_0_B8_0cubesdown_small}}}}}
  \mbox{\sidesubfloat[]{\includegraphics[scale=\scalefactor]{{{mon8_0_B8_0cubestwist_small}}}}}
  \caption{The \emph{metastable} $B=8$ Skyrmions in the miscible
    BEC-Skyrme 2+4 model.
    The three metastable solutions are: (a) two degenerate cubes, (b)
    two nondegenerate cubes, and (c) a fully degenerate fullerene-like
    solution. 
    The four columns show isosurfaces of: the topological baryon
    charge density $\mathcal{B}$, the energy density $\mathcal{E}$,
    the potential $V$, and the vorticities $\mathcal{Q}_{1,2}$ in
    $\phi_{1,2}$ (with red and blue, respectively).
    The color scheme used for the first 3 columns is described in the
    text. Each row corresponds to a different value of the potential
    parameter $M^2$.
  }
  \label{fig:B8_2+4_metastable}
  \end{center}
\end{figure}

\begin{figure}[!htp]
\begin{center}
  \renewcommand{\scalefactor}{0.204}%0.102
  \renewcommand{\scalefactorsix}{0.136}%0.068
  \mbox{\sidesubfloat[]{\includegraphics[scale=\scalefactor]{{{mon8_1_B8_1cubes_small}}}}}
  \mbox{\sidesubfloat[]{\includegraphics[scale=\scalefactor]{{{mon8_1_B8_1cubesdown2_small}}}}}
  \mbox{\sidesubfloat[]{\includegraphics[scale=\scalefactor]{{{mon8_1_B8_1cubestwist_small}}}}}
  \caption{The \emph{metastable} $B=8$ Skyrmions in the miscible
    BEC-Skyrme 2+6 model.
    The three metastable solutions are: (a) two degenerate cubes, (b)
    two nondegenerate cubes, and (c) a fully degenerate fullerene-like
    solution. 
    The four columns show isosurfaces of: the topological baryon
    charge density $\mathcal{B}$, the energy density $\mathcal{E}$,
    the potential $V$, and the vorticities $\mathcal{Q}_{1,2}$ in
    $\phi_{1,2}$ (with red and blue, respectively).
    The color scheme used for the first 3 columns is described in the
    text. Each row corresponds to a different value of the potential
    parameter $M^2$.
  }
  \label{fig:B8_2+6_metastable}
  \end{center}
\end{figure}

The story for the first seven baryon numbers (one through seven) was
that the solutions with degenerate vortices had higher energies than
solutions with nondegenerate vortices.
For the $B=8$ Skyrmion it turns out to be more complicated.
First of all, there exists both a fullerene-like solution, which is
predicted by the rational map approximation \cite{Houghton:1997kg} as
well as a solution that is made up of two cubes attached to each other
\cite{Battye:2006na}.
In the Skyrme model with a standard pion mass there are in fact two
solutions, which are made of two cubes ($B=4$ solutions) next to each
other: One is a translation of the first, whereas in the second
solution the translated copy is rotated by 90 degrees around the axis
joining them.
Both of these $B=8$ Skyrmions have lower energy than the
fullerene-like dihedrally symmetric solution, once a pion mass is
turned on \cite{Battye:2006na} and the critical value for the mass
turns out to be rather small.
In this model, on the other hand, the solutions with the two cubes are
not lower in energy for any value of the potential parameter $M$.
Furthermore, only the first version of the two cubes exist; that is,
the one where the $B=4$ solutions are translated copies of each
other, see Figs.~\ref{fig:B8_2+4_metastable}(a,b) and
\ref{fig:B8_2+6_metastable}(a,b).
By that, we mean that the twisted solution where one of the cubes is
rotated by 90 degrees around the axis joining them, does not exist in
this model as it collapses into a fullerene-like solution, see
Figs.~\ref{fig:B8_2+4_metastable}(c) and
\ref{fig:B8_2+6_metastable}(c), albeit with 
higher energy than the original $D_{6d}$ symmetric solution predicted
by the rational map approximation -- this is because this
fullerene-like solution is fully degenerate at small $M$.
This collapse of the twisted double cube $B=8$ Skyrmion has been
observed also in other Skyrme-like models, see
Ref.~\cite{Gudnason:2020arj}.
The existence of another fullerene-like solution has also been
observed in Ref.~\cite{Gudnason:2020arj}.
Of course, such existence or nonexistence of the different solutions
and which one is the global minimizer of the energy functional, is
very dependent on the model and in particular on the potential.

The energies of the different $B=8$ solutions are shown in
Fig.~\ref{fig:en58}(b)-(d).
It turns out that the $D_{6d}$ dihedrally symmetric (which is the
symmetry only at $M=0$) fullerene-like
solution is the global minimizer of the energy functional for all
values of $M$ studied in this paper -- even though it is partially
degenerate. 
The lowest-lying metastable state is another fullerene like structure,
which is fully degenerate.
The highest-energy solutions are given by two cubes joined together.
It turns out there are indeed two different $B=8$ solutions made of
cubes, but the difference is not whether they are rotated with
respect to each other or not. The difference between them is whether
the vorticities are degenerate or not. As one could guess by now, the
one with 
degenerate vorticities has the highest energy and is unstable above
$M\sim 2.1$ for both the 2+4 and the 2+6 models, see
Figs.~\ref{fig:B8_2+4_metastable}(a) and
\ref{fig:B8_2+6_metastable}(a).
A lower-lying $B=8$ solution made of $B=4$ cubes has nondegenerate
vorticities and is metastable for all the values of $M$ studied, see
Figs.~\ref{fig:B8_2+4_metastable}(b) and 
\ref{fig:B8_2+6_metastable}(b).
Although this solution may look degenerate, it is not according to
definition \ref{def:1}; the vorticities are just going very close to
one another for the red vortices $(\phi_1)$. 
The mystery remains, however, that the stable solution seems to have
partially degenerate vortices.

\section{Discussion}\label{sec:discussion}

In this paper, we have considered the miscible BEC-Skyrme model, which
is the generalized Skyrme model with fourth-order and sixth-order
derivative terms augmented by the BEC-inspired potential, but with the
opposite overall sign of the potential compared with the previously considered immiscible BEC-Skyrme model.
The symmetries of the Lagrangian are unchanged, but the vacuum state
is now connected and hence the continuous symmetry is completely broken.
The interesting point is that, unlike the immiscible BEC Skyrme model,
this model possesses two physical vortex strings: one in $\phi_1$ and
another in $\phi_2$, with $\phi_{1,2}$ being the two complex fields of
the (nonlinear sigma) model. 
These two vortex lines have been proven in
Ref.~\cite{Gudnason:2020luj} to have linking number $Q=B$ equal to the
baryon number, i.e.~the topological degree of the Skyrmion, under the
condition that a certain projection is regular.
It so happens that choosing a standard frame for the Skyrmion often
yields a situation which is not regular, and the vortex lines
degenerate -- thus making it impossible to define the linking number
(this has no consequence for the definition of the baryon number).
The potential at hand in this model makes such a degeneracy
energetically unfavorable and hence for large enough potential
parameter, $M$, the model naturally possesses Skyrmion solutions which
are made of nondegenerate vortex lines that are linked exactly $B$
times, with $B$ being the topological degree of the Skyrmions. 

The general lesson learned is that if both flavors of vortex
(i.e.~both the red and the blue vortices) are degenerate, then the
solution is a metastable state.
A lower energy solution can be found where the degeneracy is broken
in one or both flavors of vortex.
We can thus make the following conjecture:
\begin{conjecture}
  The lowest-energy Skyrmion solution in the miscible BEC-Skyrme model
  has nondegenerate vortices in both vortex species for all values of
  $M$ and for all topological degrees $B$.
\end{conjecture}
The conjecture is verified by the calculations in the topological
sectors $B=1$ through $B=7$ in this paper. 

A curious exception seems to occur for the $B=8$ topological
sector, where the lowest-energy solution we have found -- for any
value of the potential parameter $M$ -- turns out to be partially
degenerate.
That is, the red vortices (i.e.~those of $\phi_1$) are degenerate,
whereas the blue vortices (i.e.~those of $\phi_2$) are nondegenerate.
We did find a Skyrmion solution in this topological sector with
nondegenerate vortices, which is a solution made of two $B=4$ cubes
joined together.
It turned out, however, to have a higher energy than the partially
degenerate fullerene-like solution.
It would be interesting if this is an exception, or there actually
exists a lower-energy solution for the $B=8$ sector that we somehow
did not find.
As an argument in favor of this hypothesis, we found that the
difference in energy between the nondegenerate and the partially
degenerate Skyrmion solutions in the $B=4$ sector turned out to be
extremely small.
Therefore, increasing the potential parameter, $M$, may not be a
viable technique in such a situation, as incredibly large values of $M$
could be needed for finding the lowest-energy state (if such a state
exists).
Other approaches for searching for the lowest-energy state may be
needed.

It would be interesting to study the case in which a
``Josephson (or Rabi) term'' $\phi_1^*\phi_2 + {\rm c.c.}$ is added to
the potential.   
In the case of two-component BECs, this term induces a sine-Gordon
soliton stretching between two types of vortices, exhibiting vortex
confinement
\cite{Son:2001td,Kasamatsu:2004tvg,Cipriani:2013nya,Tylutki:2016mgy,Calderaro:2017,Eto:2017rfr,Kobayashi:2018ezm,Eto:2019uhe}. 
Several aspects of vortices in this case have been studied extensively,
in particular, in two spatial dimensions, such as a vortex lattice
\cite{Cipriani:2013nya}, confinement
\cite{Tylutki:2016mgy,Eto:2017rfr}, dynamics of vortices
\cite{Calderaro:2017}, collisions of vortices \cite{Eto:2019uhe}, as
well as the phase structure \cite{Kobayashi:2018ezm}.
In our case, two linked vortex lines will be connected by a
minimal-surface soliton sheet, like a soap film.

In Ref.~\cite{Gudnason:2020luj} we found that there exists a certain
projection with which a baryon contains linked vortices.
In this paper, we have shown that with a certain potential, those
linked vortices can become physical.

\subsection*{Acknowledgments}

S.B.G.~thanks the Outstanding Talent Program of Henan University for
partial support.
The work of S.B.G.~is supported by the National Natural Science
Foundation of China (Grant No.~11675223). 
The work of M.N.~is supported in part by JSPS Grant-in-Aid for
Scientific Research (KAKENHI Grant No.~18H01217).

\end{document}